%% file: main.tex
\UseRawInputEncoding
\documentclass[12pt,twoside]{mitthesis}
\usepackage{lgrind}

\usepackage{cmap}
\usepackage[T1]{fontenc}
\usepackage{amsmath}
\usepackage{amsmath}

\usepackage{tikz}

\usepackage[compat=1.1.0]{tikz-feynman}

\def\all{all}
\ifx\files\all  \else 

\usepackage{pdfpages}
\usepackage{xspace}
\usepackage{hyperref}
\newcommand{\hp}{\texorpdfstring{\ensuremath{\mathrm{H}^+}\xspace}{H+}}
\newcommand{\htp}{\texorpdfstring{\ensuremath{\mathrm{H}_2^+}\xspace}{H2+}}
\newcommand{\hthp}{\texorpdfstring{\ensuremath{\mathrm{H}_3^+}\xspace}{H3+}}
\newcommand{\hyd}{\texorpdfstring{\ensuremath{\mathrm{H}_2}\xspace}{H2}}
\usepackage{struktex}
\usepackage{tikz}
\usepackage{graphicx}
\usepackage{url}
\usepackage{fancyvrb}
\usepackage{color}
\usepackage{tikz}
\usepackage{hyperref}
\usepackage{float}

\begin{document}

\usetikzlibrary{shapes,arrows,positioning,shapes.geometric}

\newcommand{\BE}[0]{\begin{equation}}
\newcommand{\EE}[0]{\end{equation}}
\newcommand{\BEA}[0]{\begin{eqnarray}}
\newcommand{\EEA}[0]{\end{eqnarray}}

\newcommand{\ue}{\texorpdfstring{\ensuremath{\mathrm{U}^{238}}\xspace}{U238}}
\newcommand{\uf}{\texorpdfstring{\ensuremath{\mathrm{U}^{235}}\xspace}{U235}}
\newcommand{\un}{\texorpdfstring{\ensuremath{\mathrm{U}^{235}}\xspace}{U235}}

\newcommand{\qqref}[1]{Equation~\eqref{#1}}
\newcommand{\figref}[1]{FIG.~\ref{#1}}
\newcommand{\tabref}[1]{TAB.~\ref{#1}}
\newcommand{\secref}[1]{\S\ref{#1}} 
\newcommand{\chapref}[1]{Chapter~\ref{#1}}

\include{cover}

\pagestyle{plain}

\include{contents}

\include{chap1}

\include{chap2}

\include{chap3}

\include{chap4}

\include{chap5}

\include{chap6}
\include{chap7}

\include{Conclusions}
\appendix

\include{appa}
\include{appb}
\include{biblio}
\end{document}

%% file: cover.tex
\title{High Power Cyclotrons: The Bridge Between Beyond the Standard Model Physics, Computation, and Medical Applications}

\author{Loyd Waites}
\department{Department of Physics}

\degree{Doctor of Philosophy}

\degreemonth{December}
\degreeyear{2022}
\thesisdate{Dec 9, 2022}

\supervisor{Janet Marie Conrad}{Professor of Physics}

\chairman{Lindley Winslow}{Associate Department Head of Physics, MIT}

\maketitle

\cleardoublepage
\setcounter{savepage}{\thepage}
\begin{abstractpage}
\input{abstract}
\end{abstractpage}

\cleardoublepage

\section*{Acknowledgments}

This thesis is in truth the culmination of a dream I have had since I was in 2nd grade. After applying to MIT a total of four times, facing rejection, failure, and self doubt, I'm here. 

Something MIT has taught me is the difference between something that is a lot of work and something that is hard. Something that takes a lot of work can spend hours, days or weeks grinding away at an inevitable conclusion. When something is hard, you have no idea how to get there, or even if "there" exists. Sometimes there's no answer. Sometimes you have to completely change your way of thinking. When something is hard, you find yourself praying that it would just be a lot of work instead. I can grind out hours and hours of work, lock my door and not leave until it's done. But when something is hard, no matter how many hours you throw at it, it might go nowhere.

This thesis was a lot of work, but it was also hard. And when something is hard, I found I have a great group of people to turn to for help. Without their feedback, their ideas, mentorship and support this thesis would not be possible.

The first person I must thank is my advisor, Janet Conrad. My first interaction with her was a long and excited email that made me think "Wow, I might actually be able to pull this off." She was a devoted mentor in many ways, from coming in on weekends the help me prepare for exams, to sending me emails late into the night. I do not think I have met someone who so deeply loves physics, and is so motivated not only by drive and grit, but by passion and excitement. 

But a research group is truly a village that raises a student. I also have to thank Daniel Winklehner for years of holding my hand through the world of accelerator physics, finding the bugs in my code, and being there on a day to day basis for every goofy, ridiculous question and problem I raised. I appreciate your patience in dealing with me, as I think you more than anyone was exposed to my foolishness. I also need to thank Jose Alonso for his guidance, career advice, and perspective. Not only did he help me understand detectors and medical isotopes, but his network is sometimes quite surprising. (Ranging from the husband of my dental secretary, to a scientist at Raytheon who offered me a job after seeing Jose was a coauthor of mine.) Larry Bartozek, who in addition to being an all around cool dude and devote Trekkie, saved my butt on my CAD models multiple times. Larry reminded me, that despite how much we may pretend to be, physicists are NOT mechanical engineers.  The group is made up of several amazing students who form a supportive network for each other along the road, helping with basic questions we're too embarrassed to go to our advisors with, to large concepts of underlying physics. Alex, Lauren, Nick, Marjon, Gabriel, Josh, Phillip, Thomas, and most of all my partner for years Joe, have all been invaluable sources of information, advice and support. In addition, outside groups like Bhaskar, Wei-chi, Doojin, and especially Adrian from Texas A and M who were a huge part of the axion paper.

When I am asked about what makes MIT special the answer I think is not entirely obvious. The reality is that the education is actually the least important part about this place. MIT OCW offers all their courses, for free, online, with all tests and homeworks (a valuable resource I myself often use even as a student). Often professors are also more than happy to help with miscellaneous emails. So to be blunt, if I can get everything online for free to study on my own time, why bother going to MIT?

The people. It's the people that are truly nowhere else.

The most significant parts of my education I don't think were in the classrooms, but in the desperate knocking on doors of my classmates. I lost count of all the times I knocked on the doors of Brandon, Field, Jeff or Chiara asking for help with some mystery piece of physics, and how that fell into conversation and friendship. Then of course if I was really lost I'd head over to the theorists in the CTP like Greg, Sam, Ryan, Eric or Shraya, and next thing I knew I was sitting in front of a chalk board filled with diagrams and equations that would suddenly make everything so clear- and make no question seem dumb. I truly think there is a certain pride that is taken in being able to explain a topic to someone, or to teach, and often this was the motivation for us. That we wanted to say "I get it! I can teach you!" and that if someone did not know it would just... bug them until they knew. It was a pride I wish I had been on the giving end, more often than the receiving end, but am so lucky to have worked with so many brilliant people who helped to get me to this point.

My friends at MIT extend past this of course- our xapiens group with Josh, Tom, Siranush, Bobby, and several more scratched my biotech itch throughout school, and even hugely altered my future career plans. I am proud that as a particle physicist I've learned so much bio, but again it truly was talking through these concepts and papers with my peers that facilitated my learning. I got to meet some amazing people through this group, and not only learned information, but how to think and work through problems in a linear and organized way. 

But the first people I met at MIT were actually who I call now my Kareoke Crew. Coming to the open house I was actually so nervous that I hid in the bathroom before going in. When I eventually got it together to go outside, a friendly Constantine was standing before me, welcoming me, and mentioning he was president of the PGSC. To which I immediately though "Oh, so he has to be nice to me..." After years of friendship with Stan, I laugh at that thought now. He introduced me to many, but I ended up crashing on the couch of two very tall, animated, and brilliant guys named Tom and Graeme (this, I might add, was not the only time this happened over the course of my grad school career.) 

Tom told me later that when he was first introduced to "professorownage@gmail.com" that he wasn't sure if I was going to be hilarious, or just very strange, and I don't recall him laughing at many of my jokes. But weeks later they invited me to the first karaoke performance of my life. At the time, my knees shook, and I barely was able to get out "Boulevard of Broken Dreams." Now I have grown to  become a bit of a karaoke diva. One of the highlights of my graduate school career was singing on stage at the Christmas party, and terrifying Janet who had no idea what I was doing. I am still a bit amazed that Bolek just gave me the OK to sing in front of everyone, asking no questions if I had any kind of musical experience of any kind. 

Camille, Ola, Peter, Ryan, Jenny, Carmen, and the very talented Adam joined our team as the years went on. We became a bit of an unofficial club that religiously attended karaoke every Thursday. I would practice in my car all week, plan out our duets, we got a bit too into it. And loved every minute of it. 

MIT would not be complete without the wonderful administrative staff that was just as much of the experience. Cathy and Sydney were hugely supportive going truly above and beyond in my time in graduate school. Not only responsible and fantastic administrators, but personable, kind, and going to go beyond the call of duty without ever being asked.

The facilities staff were the lynch pin of MIT and building 26. From moving tables, cranes, and even borrowing the LNS van, Jack, Joe, and of course the late Lee always had my back.

On the home front, I of course am grateful to my parents for creating me that day in that testube. I hope to think that I am a pretty successful science experiment. Or perhaps a genetic mishap gone terribly wrong, bound to roam the Transylvanian countryside terrorizing villagers and tossing flowers in a pond until I am chased away by torches and pitchforks. But unlike Dr. Frankenstein to his creation, my parents have been hugely supportive in my dreams for MIT. Despite not being exactly sure what it is that I do, they have still pushed me, and I have been blessed that I never wondered if my parents were proud of me. 

But my family is more than my parents- my uncle Arty, not related in any way, has been more a part of my family than anyone I share blood with. Whenever it hit the fan, and believe me during these past few years it certainly did, Arty has been the one single person I can undoubtedly rely on, and do so without any request for reciprocation. I owe him a debt I cannot repay, but can only appreciate. 

The other part of my family is made up of the friends I have had for quite literally over 20 years. While they weren't exactly helping me with my homework, they gave me a strange gift that I think put me at an advantage over many MIT students; perspective. I saw many of my classmates undergo existential crises, some undergo full breakdowns, but I was blessed with a group who, frankly, didn't care. And it was the best thing in the world. Stress from work disappears, as does any foolish pride. I could win the Nobel prize and the next day I'd hear "Nobel prize? More like dust-del prize. Go eat some brownies, Loyd." I'm a rocket scientist, a astrophysicist, curing cancer, building nukes to protect our freedom, whatever the flavor of the day is. But really it doesn't matter, because I'm really the kid who played aliens with Liam on the eagle's nest in kindergarten, and football with Cory Squared and Chris for 10 years, and the nerdy vegetarian who gets falafel in his tacos on Tuesday. I will say I was impressed by Cory, a now an excellent football coach, once reciting my research topic really well at a bar in front of a bunch of people I just met. 
I also need to thank Janel for being there through all the ups and downs of the wild ride. She has supported me through some of the most stressful times of my life, and grounded me when I've been blindsided. Her patience is truly unparalleled. She is a fantastic partner, and a friend.

The last person I would like to thank may not be the most obvious, but is one of the most important. When looking at my career, there is a clear inflection point. There is a single day when suddenly everything changed. That is the day I met JP Laine. JP saw something in me that I did not see in myself. When I met JP, I had no serious job offers, I was doing poorly in my stat-mech class (A subject that would come to haunt me throughout my physics career), and was worried I was not even graduate. I was losing faith in my ability to be a physicist. But I LOVED Draper. When I walked in, I felt like I was in the Avengers. When I finally sat down with JP, I had no idea what to expect. I handed him my resume, he looked it over and nodded.
 
 Five minutes into the interview I already had an answer. 
 
"Ah yes. Fantastic." He nodded again. "I can't say this for you know.... legal reasons and what not, but between you and me, you have the job."

I was totally stunned. 

"Do you have any questions?"

"Yes," I thought. "Are you sure?"

But I had to think of something, and think of it fast. I couldn't just say I had no questions after being offered my dream job on the spot. 
"So I heard about this Draper fellow program..."

"Ah yes. You will work here for a bit, then you will be a Draper fellow and we will fund you through grad school. We will send you to MIT on full scholarship, and everything will be fantastic."

Oh I get it, this guy was out of his mind. 

But he wasn't. Somehow, he knew. Somehow everything JP told me came true. And since that day, my career as a scientist has never been the same.

%% file: abstract.tex
%
%
%
The IsoDAR cyclotron is a 60 MeV cyclotron designed to output 10mA of protons in order to be a driver for a neutrino experiment. Coupling the high flux generated by the IsoDAR system with a kiloton neutrino detector will provide sterile neutrino exclusion searches covering anomalous regions indicated by short baseline experiments. Simultaneously, the coupling of a high power target and kiloton detector allows for the investigation of dark matter candidates, namely axion-like particles. We have shown that nuclear excitations within the IsoDAR target create a unique opportunity to produce axions and detect monoenergetic peaks with the nearby kiloton detector.  Beyond this, the high power produced by the IsoDAR cyclotron can be used for applications beyond particle physics. The IsoDAR cyclotron accelerates and extracts H$_2^+$, which allows the beam to be split  downstream, a versatile and important development to alleviate the problem of producing high-power targets for the medical isotope community.    This thesis presents a proposal for production of an more than an order of magnitude higher rates than are available at present for certain highly-need medical isotopes, including Ac-225.

In developing the proof of principle of this state-of-the-art cyclotron, the results in this thesis focus on two points related to the production and transport of ions to the accelerator. The first step was to construct an H$_2^+$ ion source with the necessary excellent emittance parameters and low contamination of non-H$_2^+$ ions.   In this thesis, we report results from a multi-cusp ion source that meets these requirements and produces a record level of high purity, low emittance H$_2^+$ current.
The second has been to design a radio-frequency quadrupole (RFQ) that will allow for gentle bunching of the high current before injection.  This is the first use of axial direct injection with a compact cyclotron.    This thesis reports the first application of machine learning to the RFQ design.   These tools enable the high currents required by IsoDAR cyclotron, leading to  important impact on the accelerator, medical, and physics communities.

%% file: contents.tex
\tableofcontents
\newpage
\listoffigures
\newpage
\listoftables

%% file: chap1.tex
\chapter{Introduction}

\section{Accelerators and Their Applications}

When a man builds a bridge it is the wood that holds it together. Years of work can be done piecing together the building blocks in just the right way to make the largest bridge, to construct a marvel of perfectly optimized engineering. But when a new material is made, when steel replaces wood, the paradigm shifts. The bridges are made larger, the buildings taller, the tools tougher. New technology that was impossible before is made. The automobile, the internal combustion engine, the magnet. All of these things were made possible because of the tools that were made, because the very foundations of the field were pushed forward.

When someone thinks of radiation therapy, they likely think of a doctor in a lab coat, speaking to a patient with cancer. When someone thinks of particle physics, they likely think of a guy with crazy hair standing in front of a black board waving his hands yelling "Don't you see?" When someone thinks of dark matter, they likely think of the endless void of space, and the mysterious ether that seems to fill it. But there is a common thread to each of these fields, and it is hidden in the tools they use:

Accelerators.

Accelerators are the backbone of nuclear medicine and particle physics experiments. In order to push the limits of engineering and understanding, every experiment wants to turn the knobs in the same direction: more power, less cost. More protons on target, smaller accelerator.

But while many experimentalists look at accelerators as a black box, behind the curtain is multitude of information. The study of this black box is a field in itself, the study of wood, of steel, of accelerator physics. 

Finding new particles in physics requires high statistics to have proper significance. This means that we must make accelerators with a high flux to cause the maximum number of nuclear reactions and collisions. However, this comes with its own set of challenges. Increasing the beam current leads to limitations like space charge, as the beam pushes itself apart by Coulomb repulsion. These challenges must be addressed.

In medicine, a patient goes to a doctor, gets an injection, and then is scanned. That injection was a radioactive isotope, which is produced by an accelerator. Once inside the patient's body, the isotope travels the blood stream until it reaches a tumor where it decays, producing a positron (for PET imaging) a photon (for SPECT imaging) or a form of ionizing radiation such as alpha particles for therapy. Unfortunately, not all patients can afford these treatments. Many of these treatments are hugely expensive because the radioisotopes are not produced in sufficient quantities. To produce higher quantities, more protons on target must be used, and new higher power targets must designed.

This thesis is about turning the knob in the right direction. The IsoDAR experiment is based in the design of a high power cyclotron which pushes the limits of many fields, raising them all together as we make a new black box that has the potential to change physics and change lives.

\section{Producing and Observing Neutrinos} 

Neutrinos are fundamental constituents of the standard model. In the standard model there are three flavors of neutrinos, each corresponding to one of the charged leptons (electron, muon, tau). These particles have no charge but interact via the weak force, and thus have small interaction cross sections and are difficult to detect.

Neutrinos are produced in various nuclear reactions, such as proton-proton chain in the sun, but experimentally can often be associated with radioactive decays. Radioactive isotopes or particles such as pions can be produced using accelerators that irradiate a target. In the case of LSND and Miniboone, a linear accelerator used an 800 MeV proton beam on target to produce pions, which decayed producing neutrinos. IsoDAR on the other hand, will be a decay at rest experiment in which a lower, 60 MeV proton beam will collide with a beryllium target, producing neutrons, which are then absorbed by a $^7$Li sleeve. The $^8$Li beta decays, creating an isotropic spectrum of neutrinos.

One large advantage of accelerator driven systems like these as compared to solar or atmospheric neutrino experiments is their ability to be turned on and off. This provides a strong comparison of signal to background. These systems also do not rely on astrophysical observations, which are known to be model-dependent.

\begin{figure}
        \center{\includegraphics[width=.7\textwidth]
         {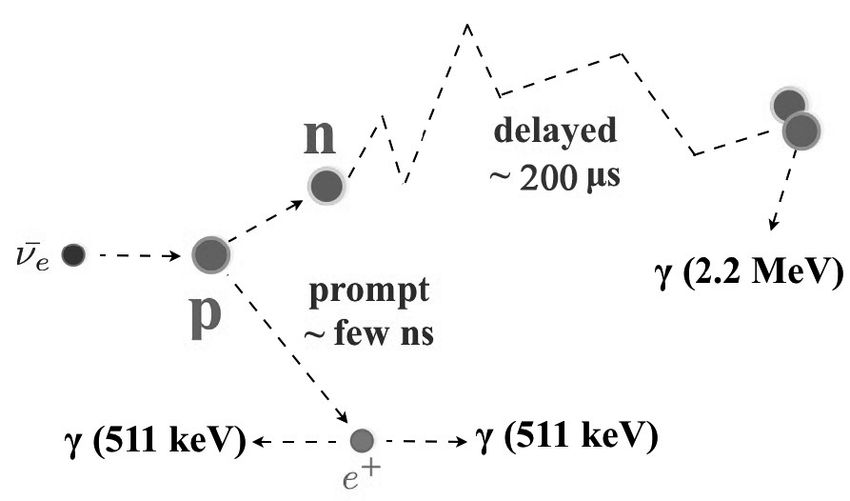}
         }
        \caption{\label{fig:ibd} Cartoon describing the process of inverse beta decay inside the oil. Taken from \cite{li2017design}. }
\end{figure}

The IsoDAR at Yemilab experiment will detect anti-electron neutrinos via inverse beta decay. This occurs when an anti-electron neutrino charge current scatters off a proton, resulting in a positron and a neutron. The positron annihilates with electrons in the fluid, creating two 511 KeV photons. These photons cause a shower, creating detectable scintillation light. The neutron recombines with a proton forming a deuteron, and emitting a 2.2 MeV photon. The 2.2 MeV photon then also showers producing a shower and detectable scintillation. There is an $\sim$200 $\mu$s delay between these two events. This provides a very clear, well understood  signal. This process is illustrated in  \figref{fig:ibd}.

In order to maximize the probability of this occurring within a detector, a high number of free protons are required. This is why oils and other hydrocarbons are frequently used in neutrino detectors, as they are relatively inexpensive and have long chains of free protons. 

It is possible to miss one of the events in the signal if recombination does not occur in the desired time frame. However, this process is well understood, and is taken into account in background calculations.

\section{Bridging the Gap: Neutrinos as Messengers for Beyond Standard Model Physics}

Dirac spinors describe quarks and massive leptons as having four possible variations: Each particle may either be left or right handed, and each particle may be matter or antimatter. It is the presence of both the left and the right handed states that gives a particle it's Dirac mass.

\begin{equation}
M_{Dirac}= M_{fermion}[\bar{\psi}_{right}\psi_{left} + \bar{\psi}_{left}\psi_{right}]
\end{equation}

It is through this mechanism that all the charged leptons are quarks have mass. 

The standard model Lagrangian also describes the weak force using the group $SU(2)_{L}$. This is because the work force only interacts with left handed chiral particles.

In addition, the standard model is broken into three generations, or flavors. This determines the lepton number or baryon number of a particle. As far as has been observed, the difference between baryon and lepton number in any interaction (B-L) is always conserved. 

There is a powerful caveat to this- a particles flavor eigenstate and mass eigenstate are not simultaneous. A single mass eigenstate exists as a superposition of flavor eigenstates. And while the mass eigenstates commute with the Hamiltonian, the flavor eigenstates do not. This means that as particles travel they travel in the Hamiltonian eigenstate, and travel as a superposition of flavor eigenstates. This leads to the property known as flavor oscillations. 

While flavor oscillations are negligible in charged particles in which the electric force dominates any event rate, they are hugely important in neutral particles that interact weakly. Namely, neutrinos.

In 2015 the Nobel Prize was awarded for the discovery of neutrino oscillations. It was found that while neutrinos could be produced in one weak eigenstate through a nuclear decay, it is possible for them to be detected in a different eigenstate at some various distance.  The superposition of neutrinos is described by the PMNS matrix:

\begin{equation}
\begin{pmatrix}
U_{e1} & U_{e2} & U_{e3} \\
U_{\mu 1} & U_{\mu 2} & U_{\mu 3} \\
U_{\tau 1} & U_{\tau 2} & U_{\tau 3} \\
\end{pmatrix}
\end{equation}

Which can be described using the oscillation equation:
\begin{equation}
   P_{\nu_{\alpha} \rightarrow \nu_{\beta}} = \delta_{\alpha \beta}-4\Sigma_{j>i} U_{\alpha i}U_{\beta i}U^{*}_{\alpha j}U^{*}_{\beta j}
   \sin^{2}(1.27 \Delta m^2_{ij} \frac{L}{E} )
\end{equation}

While the discovery of neutrino oscillations was a important result for neutrinos, it is also the first result that demonstrates there is physics beyond the standard model. While the standard model had described neutrinos as massless, if oscillations had occurred, neutrinos must travel in a superposition of flavor eigenstates, and therefore have mass. 

However this leads to more questions than answers. Why is the mass of neutrino so small? It is possible that the coupling of the neutrino to the Higgs field is simply very low, but this results in a counter-intuitive fine tuning problem.

While the obvious question that follows is "what is this mass?" We can also ask a far more bothersome question. If neutrinos have mass, does that mean they must also have a right handed wave function? One that is both neutral, and does not interact weakly? One that would make these neutrinos undetectable, or sterile?

Because neutrinos are neutral fundamental particles, they are in a unique position to have a different mass mechanism than the charged leptons. This is called the Majoranna mass. In the case of a Majoranna particle, a particle is it's own anti-particle, leading to processes such as neutrino-less double beta decay. However, the Majoranna Lagrangian requires a right-handed wave function component, and thus a right-handed neutrino \cite{Thomson:2013zua}. \begin{equation}
L_{M}= -1/2 M (\bar{\psi}_{right}^{c}\psi_{right} + {\psi}_{right}\psi_{right}^{c}]    /
\end{equation}

This is what gave rise to the theory of the see-saw mechanism, in which the neutrino masses are both Dirac and Majoranna. This simultaneously explains the small masses of neutrinos, their mass, and returns the standard model symmetry in which all particles have left and right handed counterparts. It also expands the PMNS matrix, showing that there are more mass states that may be detected\cite{de2005seesaw, branco2020type,de2007neutrino}.

The validation of these well supported mass mechanism thus requires the existence of the sterile neutrino. And while there is no definitive detection of these new mass states, there are many hints.

\begin{figure}[!htb]
        \center{\includegraphics[width=\textwidth]
         {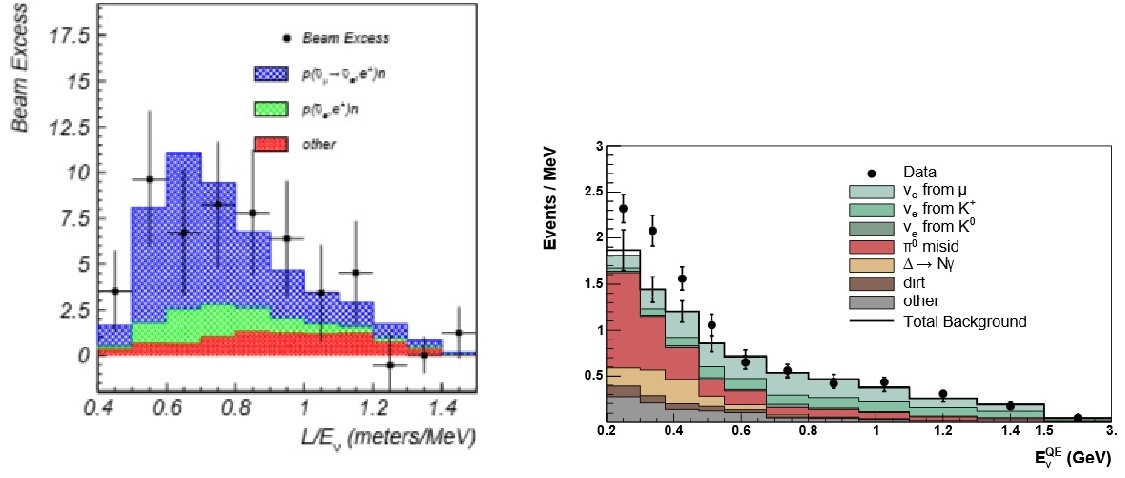}
         }
        \caption{\label{fig:LSND_miniboone} The event excess seen in LSND and Miniboone, and how a sterile neutrino fit has a stronger correlation to data than no fit. Taken from \cite{hagebout2014beyond}. \cite{inoue2004reactor} }
      \end{figure}

First found in 2001 by LSND, then later confirmed by Miniboone, event excesses lead to a stronger fit to the data in which a sterile neutrino model is used than is not, as seen in FIG \ref{fig:LSND_miniboone}.

Anomalies continue to abound and follow this same pattern, including event deficits in measurements from nuclear reactors, as well as radioactive source experiments. 

The problem facing these experiments is multifaceted. First, the neutrino cross section is very small, leading to generating sufficient statistics for definitive results to be very difficult. Furthermore, the ability to evaluate different models with different sterile neutrino properties is virtually impossible. Second, in the case of nuclear reactors, the neutrino flux does not follow the expected models, which is seen in the 5 MeV bump. 

IsoDAR looks to solve these problems by producing a high flux of anti-electron neutrinos with a well understood beta decay spectrum. Within five years of running, the IsoDAR experiment will be able to exclude all anomalous regions at five$\sigma$ level leading to a definitive evaluation of the anomalies. Not only this, but the large detector volume that IsoDAR is coupled with allows it to evaluate model sterile neutrino models, including 3+1, 3+2, and 3+1+decay.

\section{Cleaning up the Strong CP Problem: Axions}

Because neutrino experiments are constructed to search for weakly interacting particles, their designs involve powerful accelerators and large detectors. This opens the door to searches for other rare particles. A key topic of this thesis will be the application of IsoDAR to searching for the QCD axion.    

We will describe the motivation in detail later, but as an introduction I will describe the basics of axion physics and current experiments. 

\subsection{Review of Axion Experiments and Physics}
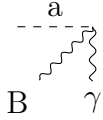
\begin{figure}[!tb]
\centering
\label{axion_bfield}

\feynmandiagram [horizontal=a to b] {
  
  a -- [scalar, edge label=\($a$\),] b,
  f1 [particle=\($B$\)] -- [photon] b -- [photon] f2 [particle=\(\gamma\) ],
};

     \caption{Feynman diagram in which an axion can can produce a photon in a magnetic field. Or, by rotation, a photon can produce axions in a magnetic field.}

\end{figure}

Axions are a particle type that was proposed to solve the strong CP problem. Namely, the strange phenomenon that the strong force has never been observed to violate charge parity symmetry. To resolve this problem, Roberto Peccei and Helen Quinn theorized that one of the standard model inputs,  $\theta$, which is the CP violating term of the QCD lagrangian, was in fact a field. Based on current observations, the coefficient of the CP-violating term in the QCD lagrangian must be suppressed by a factor of $\Theta < 10^{-10}$. While physically possible, this result seems counter-intuitively close to zero. To resolve this fine-tuning problem, the Peccei-Quinn mechanism was proposed in 1977 \cite{Peccei:1977hh}.  A particle results when this Peccei-Quinn symmetry is spontaneously broken. This result was then independently verified by Frank Wilczek  and Steven Weinberg \cite{Wilczek:1977pj,Weinberg:1977ma}. Wilczek dubbed the particle the "Axion," after a brand of soap, because it was meant to "clean up the strong CP problem."   Later it was found that with certain properties, including a weak EM coupling, this particle would also be a compelling candidate for dark matter 
\cite{Preskill:1982cy,Abbott:1982af,Dine:1982ah,Duffy:2009ig,Marsh:2015xka,Battaglieri:2017aum}. However, there is also a large area of parameter space in which the the axion is capable of being dark matter, but does not fit into the description of solving the Strong CP problem. These have thus been dubbed "Axion-Like-Particles"  or ALPs.  To search for these particles it is possible to take advantage of their standard model couplings. The properties of the axion can be determined using two parameters: its mass, and its coupling to standard model fields. For example, its coupling to the EM field is described by $g_{a\gamma}$. Due to this coupling, when an axion is in a magnetic field it is possible for it to scatter and produced a photon. The inverse is also true, if a photon is in a magnetic field, it is possible for the photon to scatter and produce an axion. This provides the vertex seen in  \figref{axion_bfield}.

While the interaction of axions with the electromagnetic field are the most commonly used in experiment, many models describe axions that can  interact via couplings with nucleons and electrons in a similar manner. These new avenues open the potential to investigate axions using different methods. This will be described in more detail in section 3.

\section{ Connecting the fundamental to the applied: Isotope Production}

\begin{center}
"When are we ever going to use THIS?"
\end{center}

A common question any teacher (or fundamental physicist) will hear whenever attempting to explain why their work is so important.

Advancing accelerator science does not solely progress particle physics. In fact, the vast majority of cyclotrons are used to produce radioisotopes for medical treatment. While IsoDAR is primarily a neutrino experiment, it is based on advances in accelerator technology, and a rising tide lifts all boats.

Medical isotopes provide a relatively non-invasive means for diagnostic and therapeutic treatment. This is achieved by injecting a biological tracer that binds to a site of interest ( such as a neurotransmitter binding site, or a tumor.) These tracers are used as a vessel to carry a radioactive isotope, which then decays at its destination. Depending on its products, this decay can create an observable signal (such as in PET scans) or cause local cell death (such as in cancer treatment), or both. These treatments can be hugely beneficial, as these tracers can be used throughout the body attacking or identifying non-localized tumors while minimizing damage to healthy cells. 

A primary issue with these treatments is economic. There is insufficient supply, particularly domestically, of many of these isotopes for them to be widely used. These isotopes are produced at accelerator facilities whose production rates are limited by beam power and maximum target power. While groups such as DOE use linear accelerators to produce isotopes, most accelerator facilities use cyclotrons due to their small footprint and relatively low cost. The isotopes are then shipped to hospitals for use. As the isotopes are transported, a fraction of these isotopes decay, causing loses. This means the locality of these facilities to hospitals, as well as the half lives of the isotopes being used, is be crucial. There are often stories of doctors quite literally running down the hall with a radioisotope to a patient in order to limit the decay of the isotope they are about to use. 

Research to increase the yields of these isotopes is ongoing, however progress is incremental. What is proposed with the IsoDAR cyclotron would be a monumental paradigm shift. If the full current of IsoDAR was utilized to produce medical isotopes, world supplies of rare isotopes could by duplicated in a matter of hours \cite{waites2020potential}. It would also provide the perfect testbed for the research and development of high power targets, which is a crucial bottleneck in the isotope industry.

%% file: chap2.tex
\chapter{The IsoDAR experiment}

\section{Review of Short-Baseline Anomalies and Sterile neutrinos}

\subsection{Pion Decay-Chain Sources}
The most famous, and most notorious, evidence for beyond the standard model neutrinos are the anomalies seen by LSND and MiniBooNE experiments.

The first hint of BSM neutrinos was found in the the Liquid Scintillator Neutrino Detector (LSND) experiment. In this experiment, a $\sim$1 mA 800 MeV proton beam collided with a target embedded in shielding producing pions. While negatively charged pions were absorbed,  positively charged pions came to a stop and then decayed to muons and muon neutrinos.  The muons then decay to positrons, anti-muon neutrinos, and electron neutrinos. These postively charged muons decayed at rest, providing a well understood neutrino spectrum. Due to the size of the beam stop and subsequent 30 m of downstream shielding, only neutrinos from pass through to the detector. Anti-electron neutrinos were then detected via inverse beta decay in LSND, which was a cylindrical 167 metric ton mineral oil scintillating detector \cite{athanassopoulos1997evidence, abe2020search}. LSND observed an excess of 87.9 $\pm$ 22.4 $\pm$ 6.0 inverse beta decay events \cite{aguilar2001evidence}. This indicated anomalous behavior that could not be explained solely by three-neutrino oscillation models \cite{diaz2022through}.  This is often interpreted as a 3.8$\sigma$ excess in the channel of $\bar{\nu}_{\mu} \rightarrow \bar{\nu}_{e}$\cite{aguilar2001evidence}.

The MiniBooNE experiment was designed as a followup and test to the LSND experiment. The MiniBooNE  would probe a similar L/E, but using a different distance and energy. To achieve this, MiniBooNE used an 8 GeV proton beam from Fermilab’s Booster Neutrino Beam (BNB) on a
beryllium target to produce mesons which would decay in flight yielding muon neutrinos. It also used a magnetic horn to separate positive and negative mesons, allowing the experiment to run in both a neutrino and anti-neutrino mode. After the horn was a decay region in which the pions would decay to neutrinos, and a 540 m of earth shielding before the detector which would absorb all other particles \cite{stancu2003miniboone}.  MiniBooNE shows a 4.8$\sigma$ excess of $\nu_{\mu} \rightarrow \nu_{e}$ oscillation events \cite{aguilar2021updated}, as well as a 2.8$\sigma$ excess of $\bar{\nu}_{\mu} \rightarrow \bar{\nu}_{e}$, both of which correspond to oscillations of $\delta m^{2}$ of $\sim$1 eV \cite{aguilar2013improved}. 

However, in 2021 MicroBooNE, which was designed to further investigate the MiniBooNE excess, released a set of papers describing an analysis in which they performed three separate analyses spanning several types of interactions (Quasi-elastic, semi-inclusive and inclusive) \cite{abratenko2022search,microboone2022search,collaboration2022search,abratenko2022search}. Studies were performed on muon neutrino and electron neutrino samples.  The muon neutrino samples consistently agreed with the Standard Model predictions.    However, the electron neutrino samples, consistently showed deficits.   Taken at face value, this allowed MicroBooNE to set limits on an observed electron-flavor excess.   However, the deficit of electron neutrino events, can be interpreted as electron flavor disappearance (${\nu}_{e} \rightarrow {\nu}_{e}$) \cite{denton2021sterile,carloni2022convolutional,aguilar2022miniboone, arguelles2021microboone}.   This was unexpected and is under investigation.

\subsection{Radioactive Source Experiments}

The Gallex experiment was designed as to observed solar neutrinos using a gallium detector. This was tested during a calibration using radioactive sources ($^{51}$C and $^{37}$Ar). A $\sim$3$\sigma$ deficit of events was found \cite{giunti2011statistical}, that has been interpreted as a potential signal for ${\nu}_{e} \rightarrow {\nu}_{e}$ disappearance.

In 2021, the BEST experiment presented results to investigate this deficit. A A 3.414-MCi $^{51}$Cr neutrino sources was placed in the middle of two separate gallium detectors. The gallium detectors measured anti-electron neutrino events via charged current interactions. The measurements were made at two distances. The BEST experiment reported a highly significant anomaly, finding an over 5$\sigma$ deficit in the number of neutrino events \cite{barinov2021results, barinov2022search,berryman2022statistical}.   With that said, the allowed region is in parameter space that had been previously explored for the simplest sterile neutrino model.
These germanium detector experiments have the disadvantage that they only are able to count charge-current neutrino interactions, and are unable to observe the path length (beyond the specific detector) or energy of individual events. 

The results of these radioactive source experiments are consistent with a sterile neutrino with $\delta m^{2} >1~{\rm eV} ^{2}$ and most favor $\delta m^{2} \sim 10~{\rm eV} ^{2}$.

\subsection{Nuclear Reactor Anomalies}

Nuclear reactors produce a unique flux with distinct features. Anti-neutrinos produced in nuclear reactors are low energy (producing $\sim 3$ MeV events), and detectors can be in close proximity ($\sim 10$ m in the case of research reactors), making both L and E very low. Nuclear reactors also produce a very high number of neutrinos as a byproduct of nuclear reactions used to produce power. This provides a unique channel and parameter space for studying a high flux of neutrinos. 
In this section, we consider production of reactor neutrinos in detail, because issues associated with this flux inspired the proposal for the IsoDAR source.

\begin{figure}[!tb]
        \center{\includegraphics[width=.5\textwidth]
         {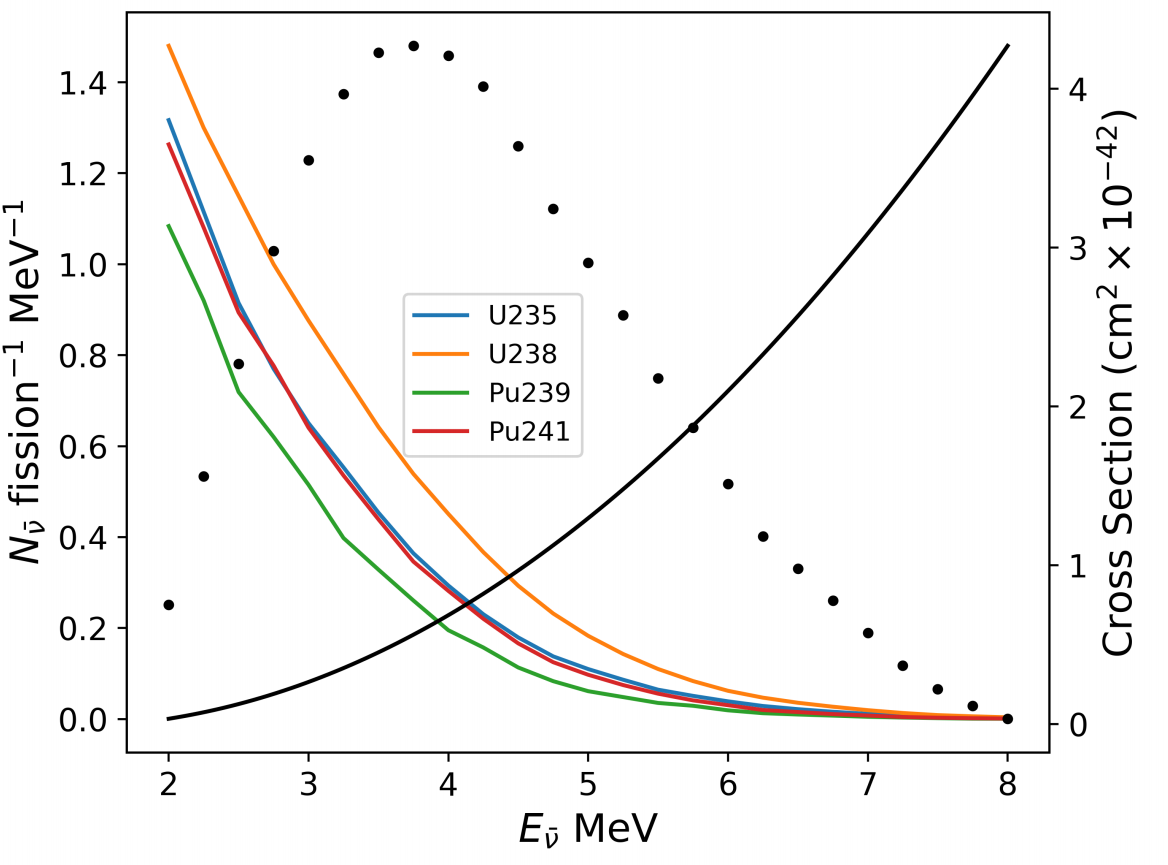}
          }
        \caption{\label{fig:reactor_flux}  Reactor neutrinos produced per fission per MeV for different materials in a nuclear reactor. In black is the inverse-beta-decay cross section. The black dots are the convolution of the two. Taken from \cite{diaz_where_2019}. }
      \end{figure}

There are four primary elements burned in a fission reactor: Uranium-235, Uranium-238, Plutonium-239, and Plutonium-241. Each of these elements produce a different yield of anti-electron neutrinos, this can be seen in \figref{fig:reactor_flux}. Therefore, when predicting the anti-electron neutrino flux from a reactor, it is important to know the relative abundance of each isotope in the reactor and how it evolves in time.

 Some fast neutrons cause \ue to fission. This creates more fission products, anti-electron neutrinos, and neutrons. A portion of fast neutrons are also captured by \ue forming \un.  \un then beta decays to Np$^{239}$, which beta decays to Pu$^{239}$, each beta decay producing more anti-electron neutrinos. The  Pu$^{239}$ continues to build up, until fast neutrons cause their fission, leading to further production of neutrons, fission products, and anti-electron neutrinos. Alternatively, thermal neutrons can be captured by Pu$^{239}$, yielding Pu$^{240}$  and Pu$^{241}$. These remain in the reactor until fast neutrons cause their fission.

Because of this,  Pu$^{239}$ and  Pu$^{241}$ production are correlated, and as more neutrinos are produced from Pu$^{239}$. More Pu$^{239}$ is produced over time, and as it increases so does the production of neutrinos. In addition. more neutrinos are produced from \uf than \ue, despite there being a higher abundance of \ue present in the fuel. This is because the production cross section of \ue is much larger.

\begin{figure}[!tb]
        \center{\includegraphics[width=\textwidth]
         {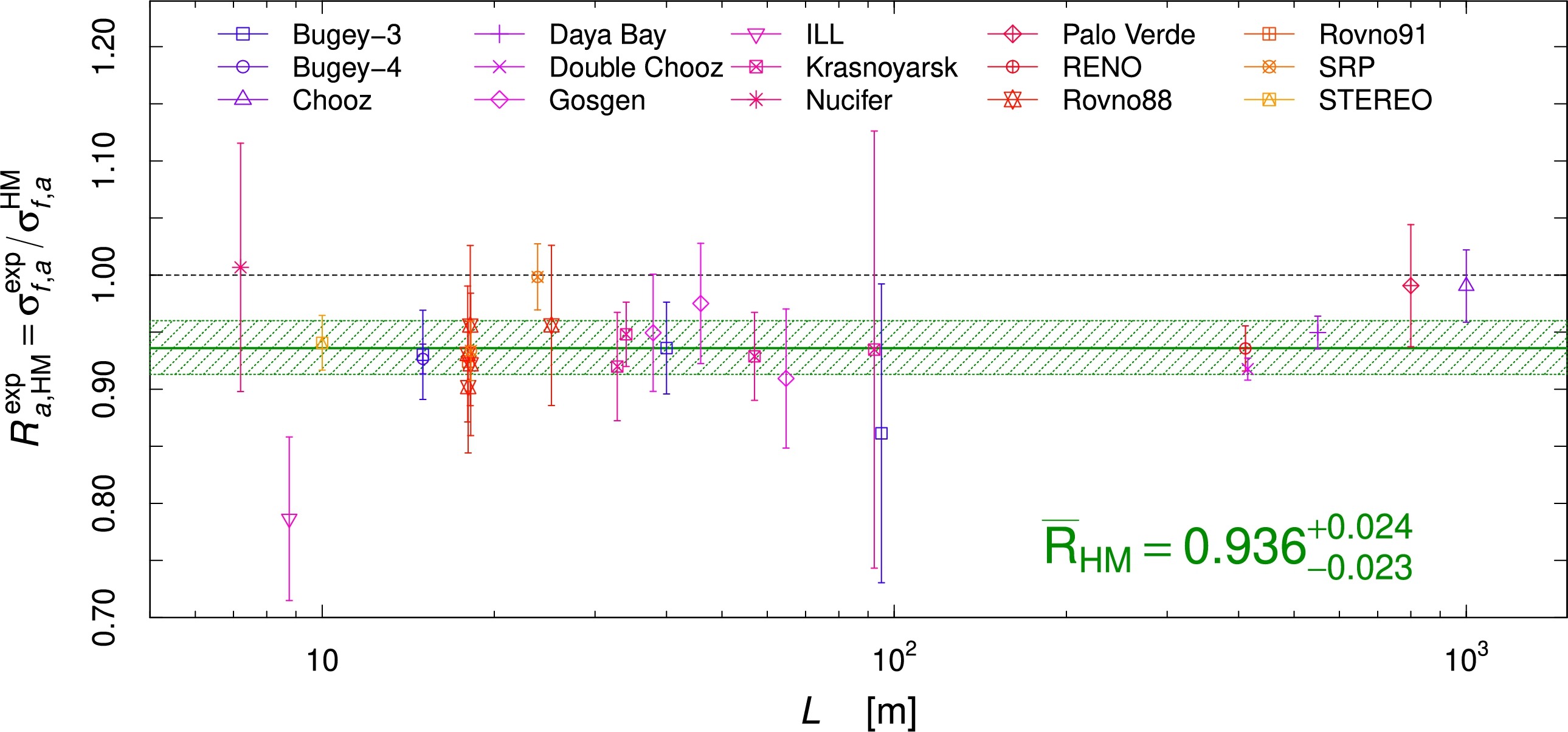}
          }
        \caption{\label{raa} Ratio of observed IBD rates as compared to the prediction made by the Huber-Mueller model, showing the deficit know as the RAA. Taken from \cite{giunti2022reactor}. }
      \end{figure}

  Reactor flux had believed to be properly modelled until 2011. When updated cross-sectional data was used to model the reactor neutrino flux, it was found that the expected reactor flux was higher than what was experimentally measured  \cite{mention2011reactor, mueller2011improved, huber2012erratum}. This is what has since been known as the reactor neutrino anomaly, or RAA. A plot describing this anomaly can be seen in \figref{raa}
  
While several models have been proposed to explain this anomaly, there is no strong consensus \cite{giunti2022reactor}.
 
 However, in this search a new anomaly has been found. In the region of 5 MeV an additional excess has been observed. This bump has been observed and confirmed by multiple reactor experiments\cite{diaz_where_2019, berryman2021sterile, huber2017neos, licciardi2021results, huber2017neos, ochoa2019daya,zacek2018evidence}, but is has not be able to be properly described by a nuclear physics or reactor model. It is regarded as a likely demonstration that the neutrino flux from a nuclear reactor is not understood. It has been proposed that additional experiments, that are not based on the uncertainties which are found in reactor experiments, should be used \cite{berryman2020reevaluating}.
 
 The PROSPECT and STEREO experiments attempted to improve these experiments by using nuclear reactors that use only \uf to simplify the experimental comparison to the neutrino flux model. However, these experiments still observed a 5 MeV bump relative to the expected Huber flux with 2.4 $\sigma$ significance.  \cite{almazan2022joint,andriamirado2021improved}
 
 While this bump in the range of 5 MeV does demonstrate problems with the currently used reactor model, it does not significantly impact the evidence for the sterile neutrino\cite{berryman2020reevaluating}. Therefore, in investigating anomalies in the search for sterile neutrinos in the anti-electron neutrino channel, it is important to have another experiment to evaluate the anomalies in this channel that are not dependent on highly variable reactor flux. 

Due to these anomalies, it is clear that the neutrino reactor flux is not well understood, and must be supplemented by additional experiments. To properly understand the source of the anomalies, having a predictable neutrino spectrum is critical. To do this, I propose producing an anti-electron neutrino flux using the IsoDAR experiment. This experiment will use a decay-at-rest source of Li$^{8}$, which produces a clear beta-decay spectrum. This will provide a direct comparison to the reactor neutrino experiments, and the much needed clarity of the anti-electron neutrino channel.

\section{Tension in 3+1 Models}

Before discussing the IsoDAR design, let us consider the combined impact of the anomalies above.    These anomalies, and experiments with no deviations from the Standard Model, are usually fit to the simplest sterile neutrino model, 3+1.

However, the modern landscape as well as \cite{diaz2022through} have shifted away from  the 3+1 sterile neutrino model to describe the anomalies found in MiniBooNE and LSND. Global fits which compare the expected sterile neutrino parameters from multiple different channels have found tension between their results. While anomalies exist in multiple different channels, the properties of the sterile neutrino to describe those anomalies do not match one another. However, as described in the previous sections, models that include more than one sterile neutrino are far more theoretically motivated than the typical 3+1 model which has been the phenomenological "go-to" for many experimentalists. This does not exclude a sterile model, it excludes the simplest sterile neutrino model.  That leaves us with a problem:   how do we know what more complex model we could explore?   Examples of more complex models include 3+1+wave packet effects, 3+1+decay and 3+2, as well as others.    Using phenomenology-inspired models and many data sets for fits has limitations. But in order to evaluate models with higher numbers of sterile neutrinos requires very high neutrino statistics which have been unavailable in the current experimental landscape.   A more desirable approach would be to introduce a source with a well-understood flux associated with a detector that can reconstruct the oscillation wave in $L/E$.   That is the IsoDAR experiment.

\section{IsoDAR Design, Motivated by the Sterile Neutrino Search}
While the results described here from many neutrino experiments are exciting, they are also quite confusing. There is no definitive measurement that is capable of differentiating different models of sterile neutrinos with high enough statistics to present convincing level of certainty. In addition, in recent years models with multiple sterile neutrinos have become the focus, but require higher statistics. This is where the IsoDAR experiment finds its niche.

Although its cyclotron has multiple applications, the IsoDAR accelerator was originally motivated to enable a definitive experiment for sterile neutrinos. Neutrino physics as a whole has faced a powerful challenge: namely, the small interaction cross section of the neutrino. There have been efforts to make more spatially efficient detectors using coherent scattering \cite{agnolet2017background, lindner2017coherent, aguilar2016results, aguilar2019exploring}, as well as massive detectors \cite{abe2018hyper,yokoyama2017hyper,seo2019neutrino,lee2020radioassay,himmel2014recent,fukuda2003super}, to try to make up for this small cross section. 

Yet many of the questions in neutrino physics remain unanswered as detectors become more and more expensive. The IsoDAR experiment turns this paradigm on its head. Rather than creating a new form of detector, this is a new form of neutrino source.  The IsoDAR source acts as a neutrino flood light in close proximity to an underground kiloton neutrino observatory, allowing for unprecedented rates from an accelerator-based experiment at energies relevant to the neutrino anomalies.

\begin{figure}
\begin{centering}
\centering
\includegraphics[width=\textwidth]{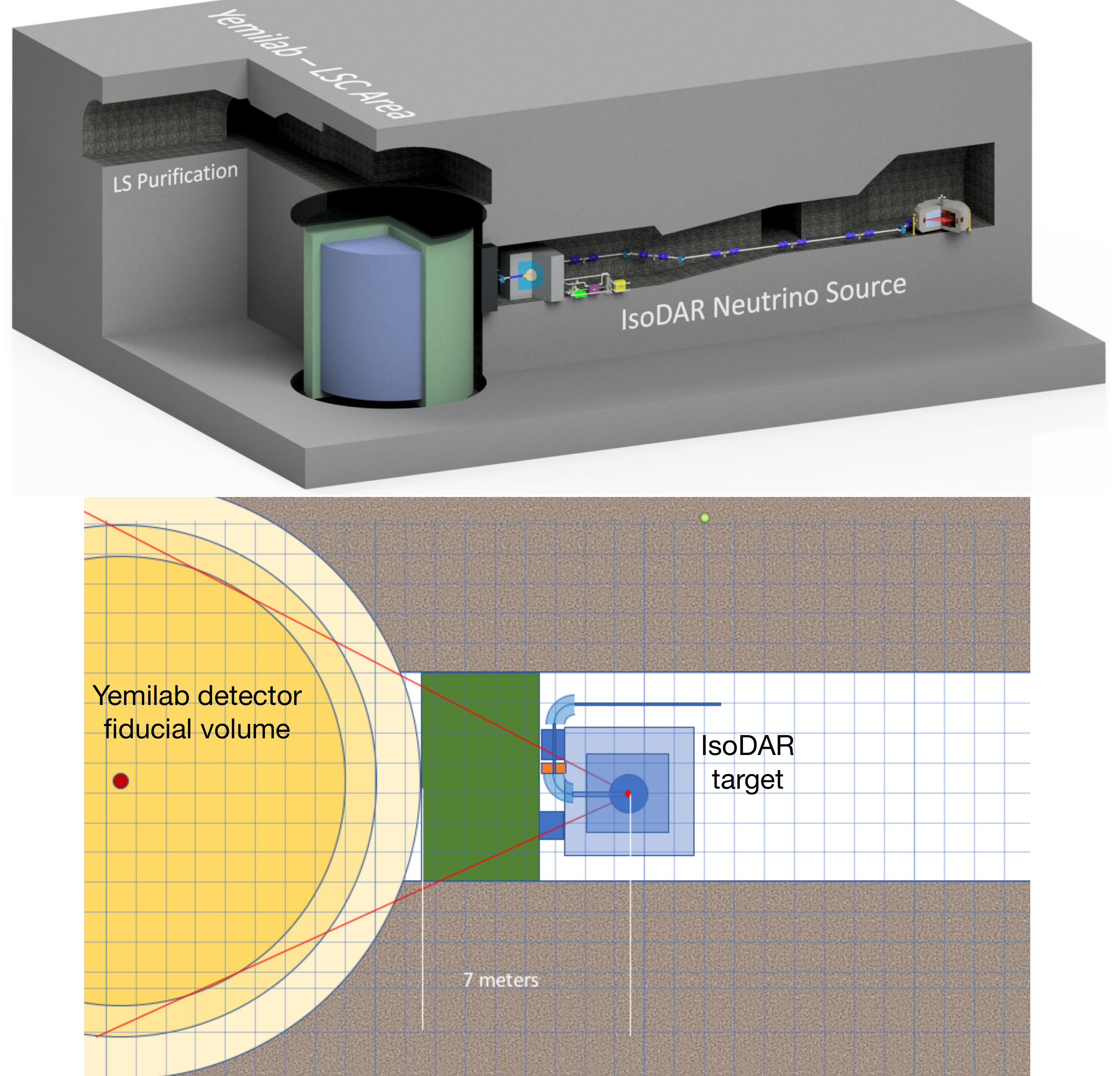}
\caption{CAD rendering and schematic of  IsoDAR@Yemilab The distance from the center of the IsoDAR target to the center of the detector is 17~m (from left-to-right in the bottom drawing, 7.5~m inner detector radius, 1.0~m buffer, 1.5~m veto, and 7~m of shielding+target). Taken from \cite{alonso2022neutrino}}
\label{yemilab_geometry}
\end{centering}
\end{figure}

The IsoDAR source is designed to use a high-power cyclotron producing a high flux of anti-electron neutrinos. The high flux of neutrinos is generated in close proximity to a kiloton neutrino detector. This is laid out schematically in \figref{yemilab_geometry}.   The source make use of  $^8$Li, which is an isotope with a half life of $\sim$800 ms.   As a result, an accelerator is required to constantly replenish the source.
The accelerator system in conjunction with the neutrino detector forms a sterile neutrino experiment, providing a 5-sigma exclusion over a large parameter space after 5 years \cite{winklehner:nima,alonso2022neutrino}. To achieve this result, the accelerator faces two major design constraints:
The accelerator is required to be built in an underground mine in close proximity to a kiloton scale neutrino detector. Therefore, the accelerator must be compact in size. This prevents use of a large separated sector cyclotron \cite{kolano2018intensity}. The accelerator must produce 10 mA of 60 MeV protons (600 kW of beam power). This is an order of magnitude higher than commercially available compact cyclotrons, typically used for medical isotope production, can deliver \cite{waites2020potential,alonso:isotope}. Several new techniques were used to address these constraints, including the acceleration of \htp and radio-frequency quadrupole (RFQ) direct injection.
This section will discuss in summary the functioning parts of IsoDAR. A much more detailed explanation can be found in the following chapter 6, appendix A and B. 

\subsection{The Accelerator}

A cyclotron producing 10 mA of 60 MeV protons is required in order to produce the needed neutrino flux for the IsoDAR experiment. There are five major subsystems: An ion source, a low energy beam transport (LEBT), a spiral inflector, an accelerating RF system, and an extraction system.

The ions for the beam are produced in a high current, low emittance ion source. In the case of IsoDAR, we begin with a beam of \htp ions. The ion source chamber is pumped with hydrogen gas where it is ionized by an  electron producing filament. The \htp is contained by using a multicusp magnetic field, and drifts through an extraction hole. Electrostatic lenses are used to extract the ions from the source and shape the beam before feeding them into the RFQ. Our ion source is explained in detail in the later chapter 4 and appendix A. 

A low energy beam transport (LEBT) is designed to take the ions from the source and bring them to the needed position in order to inject them into the cyclotron. Typically, quadrupole and dipole magnets are used to control the parameters of the beam and separate unwanted species. Depending on the the needs of the system, a LEBT is approximately 5 meters long. In the IsoDAR cyclotron, we will use a short LEBT that will inject into a radio frequency quadrupole (RFQ). The RFQ is estimated have up to a 99$\%$ transmission rate \cite{winklehner:nima}, eliminating beam losses that would occur in a LEBT, while simultaneously making the system more compact. The RFQ is specifically tuned  for a single species (in our case \htp). The different gyromagnetic ratios of each species will cause them to collide with the vanes with the RFQ, eliminating contaminating beam species.

While accelerating the beam slightly (15-70 KeV), this is not the primary function for the IsoDAR RFQ. The RFQ will act primarily as a buncher, using the radio-frequency fields to break up the beam into multiple easier to manage clusters of particles, or bunches. Each of these bunches are then packaged within a single phase window, eliminating loses due to beam acceptance into the  central region.

The RFQ injects axially into the cyclotron. This means the the direction of the beam is initially perpendicular to the plane of acceleration in the cyclotron. To align the beam into the cyclotron's plane of acceleration, the RFQ injects beam into a spiral inflector.  The spiral inflector is two electrostatic plates that steer the beam around an angle into the acceleration plane with minimal loss of energy and current. 

Once in the accelerating plane, the beam is accelerated when crossing the RF gaps within the cyclotron. The magnetic fields generated by the magnets and coils within the cyclotron cause the beam to follow a circular orbit, and repeat to cross each RF gap. As the beam gains energy, the orbit radius increases, causing the beam to spiral outward towards extraction. The cyclotron is approximately symmetric about the accelerating plane. On each side, there are four sections, or "D's" that produce the magnetic field, each separated by RF gaps. This creates an acceleration effect by using "hills and "valleys." 

The required high current calls for vertical focusing forces as the beam progresses, causing the orbit frequency to be constant throughout the cyclotron. A cyclotron with constant frequency like this is called "isochronous." This will require a radially increasing magnetic field, which is established by changing the shape of the steel (in a process called "shimming") and the coils as they extend radially. There are several ways to change the field based on the shape of the steel. For example, the vertical gaps between the two halves of the cyclotron can be decreased radially to increase the peak magnetic field. However, to maintain isochronicity only the average magnetic field around any loop must be considered. An easier way to maintain this is to extend the edges of the "hills" azimuthally so that the particles are exposed to a higher field for a greater period of time. 

 Having a radially increasing magnetic field also allows for a single-frequency RF system to be used. This eliminates the need to increase frequency as the beam progresses in order to maintain isochronicity. Having a radially increasing field rather than an increasing RF frequency allows for a continuous bunched beam, as opposing to accelerating individual bunches like with a synchro-cyclotron. This will allow for much higher average beam currents.
 
 At the last turn, the cyclotron has accelerated approximately 5 mA of \htp to 60 MeV, and the beam must be extracted from the cyclotron. On the final turn in the cyclotron, the beam will be led towards an electrostatic channel with a thin grounded septum. The high voltage on the channel will cause the beam to bend slightly, causing it to exit the cyclotron tangentially.

Once out of the electrostatic channel, the beam will enter a magnetic channel. This will serve to dampen the fields from the cyclotron to prevent them interfering with the extracted beam, as well as shape the beam to best fit the desired exit parameters. The magnetic channel is followed by focusing magnets which will control the beam after it exits the cyclotron. At this point, the \htp is stripped of an electron, yielding approximately 10 mA of protons.

Turn separation is an important factor in determining the details of extraction. Maximizing the RF voltage will also maximize the energy gain per turn, and therefore the turn separation. However, there will be some overlap in the beam between turns. This overlap will collide with the thin septum, causing damage and impeding the extraction. Based on our cyclotron simulations, it is possible to keep the power on septum below acceptable levels \cite{alonso:isodar_cdr2,alonso2022neutrino}. To further protect from damage, a stripper foil will be placed before the septum, stripping the \htp to protons and changing their gyromagnetic ratio. This will cause the protons to bend off in the field of the cyclotron, and be ejected radially. This approximately 50 $\mu$A beam of protons can be dumped, or aimed at another target in order to produce isotopes. This highlights an additional advantage of \htp beams within a cyclotron, as this safety measure would not be possible if the beam were extracting protons.

\subsection{The Power on the IsoDAR Target and Its Implications for Medical Targets}

To produce neutrinos the proton beam drives a decay at rest target. Neutrons are generated in the target. These neutrons then interact with the Lithium and Beryllium mixture in the surrounding sleeve, producing a high flux of $\bar{\nu_e}$ which can be used for searches for physics beyond the standard model. The entire system is then surrounded by steel and concrete to provide neutron shielding for the IBD detector. 

As can be see in \figref{yemilab_geometry}, the beam points in the opposite direction of the detector. This is because the neutrino production via beta decay is isotropic, but fast neutrons favor the beam direction. Therefore, facing the beam away from the detector prevents neutron detection while having no impact on the neutrino flux.

The high beam power on target required extensive thermal investigation to ensure the target's survival over the multi-year IsoDAR run time. The $^{9}$Be target is separated into three hemispheres and is cooled with a series of water pipes using heavy water. Heavy water is used  primarily due to its lower neutron absorption cross section that leads to substantially higher $^{8}$Li yields.   It has the added benefit that neutrons can be knocked out, increasing the neutron flux in the lithium sleeve, but because the cooling water is a small percentage of the target, this is not a large enhancement overall.   The quantitative advantage of heavy water over light water is $\sim$40\% improvement in  $^{8}$Li yield over water in our GEANT4 simulations \cite{alonso2021isodar}. 

This system is separated mechanically from the rest of the target. Due to its shape, this section is known as "the torpedo," and can be removed from the sleeve and replaced \cite{alonso2021isodar}. Due to the high radioactivity of the torpedo infrastructure must be put in place so that this process may be completed remotely. 

While the target has been designed at a conceptual level to handle the 600 kW of power from the IsoDAR cyclotron and produce the needed neutrino flux, optimization of this design is ongoing \cite{alonso2021isodar}.   An important point is that the target face is large (20 cm diameter) and the beam is ``painted'' (i.e. moved) across the face during running.  This reduced the thermal issues substantially. The thermal stresses to the beryllium target are being investigated by a team at Columbia University. The studies done for the IsoDAR neutrino target can be related to a similar problem in the medical and industrial sector, namely handling the thermal stresses on the targets for medical isotope production.

In order to generate medical isotopes the accelerating process is largely the same, but a different target must be used. The material of the target determines the isotope produced. To produce $^{68}$Ge (a valuable isotope used in PET scans,) a natural gallium target can be used. $^{68}$Ge is produced by a (p,2n) reaction from $^{69}$Ga. To produce $^{225}$Ac (another valuable medical isotope used in cancer treatment) would require a natural thorium target.

A driving question behind targets used for isotope production will be the amount of beam power that will be used to produce isotopes. In the case of the IsoDAR neutrino experiment, in which approximately 50 $\mu$A of current (3 KW) will be broken off to protect the septum, it is unlikely thermal stresses will be an issue. However, if we wish to increase our yield of isotopes, we may wish to create a second independent cyclotron used solely for their production. This would generate 600 KW of protons, which is a higher power than modern medical targets are capable of handling  \cite{alonso:isotopes,waites2020potential}.

There are two ways to approach this problem. Either to develop higher power targets that would be able to high handle the power of the cyclotron. Or, we can develop ways to split the beam into lower power beams that will each individually collide with a modern target.

High energy targets have been designed to handle up 50 KW,\cite{alonso:isotopes,waites2020potential}  however this has not been demonstrated and is still a factor 12 lower than required to use the full power the IsoDAR cyclotron. Given our groups expertise in accelerator and beam technology, we are developing several ways to approach method 2. 

First, within the cyclotron it is possible to place additional extraction/stripping sites. By stripping part of the beam at several places within the cyclotron the gyromagnetic frequency of the particles change, causing them to be ejected radially from the cyclotron. This will allow for four different extraction ports, although it is possible only two may be able to be used at one time \cite{alonso:isotopes,waites2020potential}. This will lessen the power requirements of the target by a factor of between 2 and 4.

It may also be possible to further split the beam using a mixture of dipole magnets and electrostatic lenses. Splitting the beam will allow for several stations that can be monitored individually for production of different isotopes or experiments. A dipole magnet with a mobile stripping foil will provide variable power to each of these stations. This provides the perfect testbed for research and development of higher power targets, which are crucial for the isotope industry.

\subsection{The IsoDAR Sleeve}

The beam strikes $^{9}$Be, which produces neutrons. The beryllium target is surrounded by an 84 cm spherical sleeve of $^{7}$Li and $^{9}$Be. The $^{7}$Li in the sleeve undergoes neutron capture, producing $^{8}$Li. The additional $^{9}$Be that is within the sleeve acts a neutron multiplier, increasing the neutron flux, and thus increasing the total production of $^{8}$Li in the sleeve. Optimization led to the choice of 75\% $^{9}$Be by mass \cite{alonso2021isodar}. $^{8}$Li undergoes $\beta$ decay, 

The purity of the $^{7}$Li is paramount to the experiment. Contamination of $^{6}$Li can lead to significantly lower production of $^{8}$Li because the neutron capture cross section of $^{6}$Li is three orders of magnitude higher than that of $^{7}$Li. Therefore, in designing this experiment we have assumed a 99.99\% pure $^{7}$Li sleeve. However, if the sleeve is even 99.995\% pure $^{7}$Li, which we are told is possible in fabrication, it would lead to a significant increase in neutrino production. 

In addition this sleeve has been designed to be highly compact, in order to have a precise knowledge of $L$ when calculating the neutrino $L/E$. The neutron shielding outside of the sleeve is determined by regulatory limits, and uses boron-rich concrete in order to maximally capture neutrons in the smallest space.

\subsection{Sensitivities and Projections for Sterile Neutrinos}
\begin{figure}[!ptb]
\begin{centering}

\includegraphics[width=\textwidth]{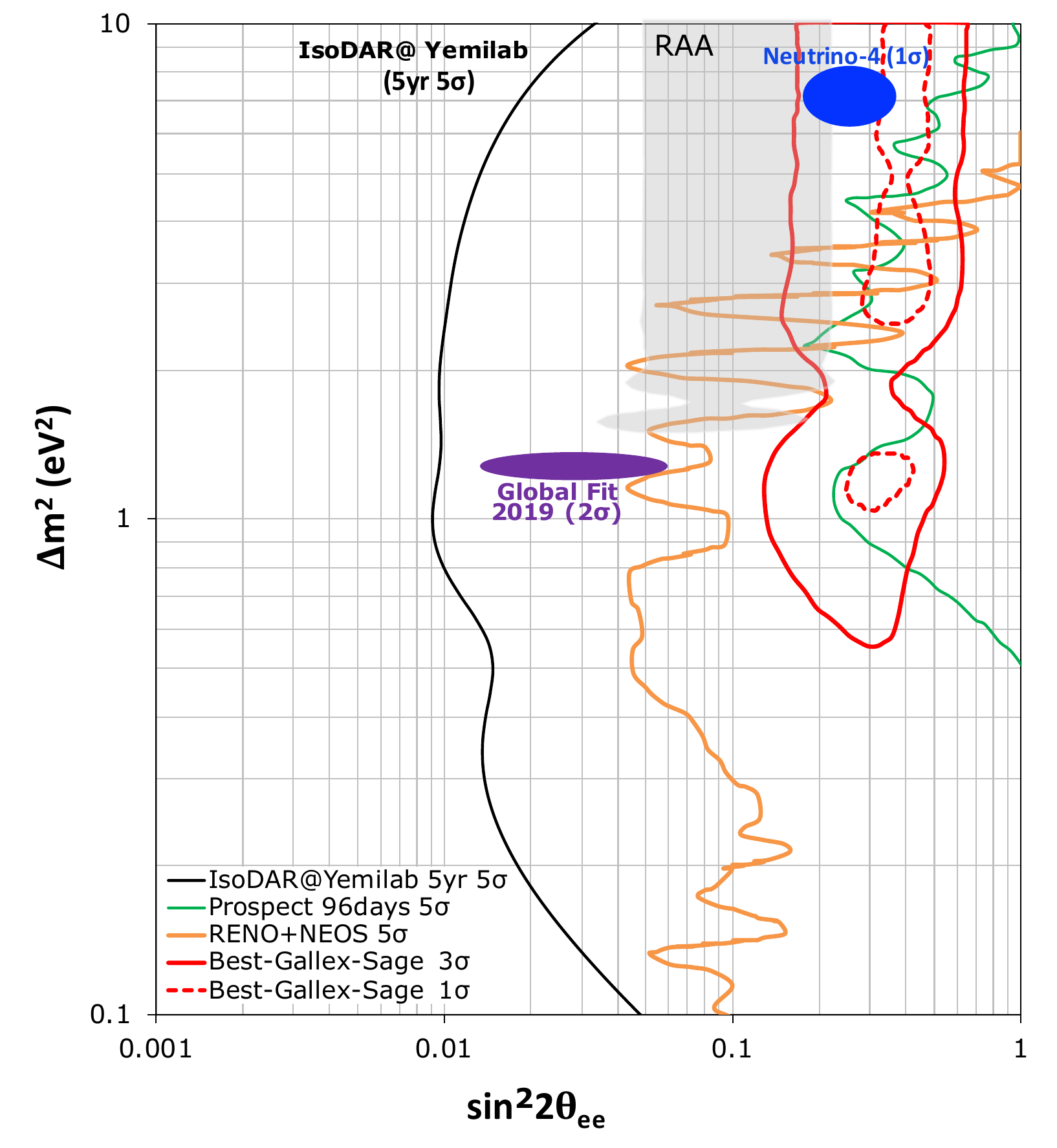}

\caption{The 5$\sigma$ sensitivity achievable by the IsoDAR@Yemilab experiment in 5 years of running, compared to a number of existing electron-flavor disappearance measurements. Taken from \cite{alonso2022neutrino}}
\label{sensitivity}
\end{centering}
\end{figure}

\begin{figure}[tb]       
\begin{center}
\includegraphics[width=\textwidth]{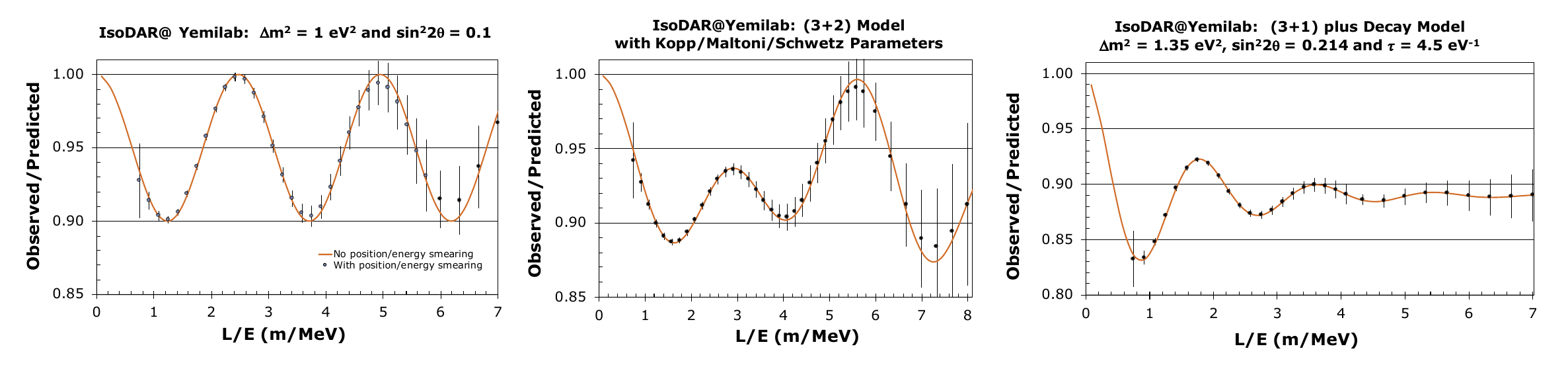}
\end{center}  

\caption{\label{wiggles} The IsoDAR@Yemilab capability to measure oscillations under three example representative new physics scenarios: a 3+1 model (left), a 3+2 model (center), and a 3+1 with neutrino decay model consistent with the 95\% allowed region observed at IceCube (right)~\cite{Moulai:2021zey}. The points on the left and middle plots include position and energy smearing based on the expected Yemilab detector resolutions. The plot on the right does not include this smearing. Taken from \cite{alonso2022neutrino}.}
\end{figure}

Including detection efficiency, the IsoDAR experiment running at Yemilab is expected to produce $1.67 \times 10^{6}$ IBD events in 5 years of running. This includes the estimated IsoDAR up-time that is 80\%, primarily to account for maintenance.  This allows world leading sensitivity to previously unexplored regions of sterile neutrino parameter space and other physics models. The IsoDAR experiment can then be compared to the experiments listed previously in this chapter, as well as global fits. This limit can be see in \figref{sensitivity}

In addition, IsoDAR's high statistics allow for modelless tests due to ability to reconstruct the oscillation wave.  This gives information for new models. In addition to the typical 3 + 1 model, IsoDAR can investigate a model with two sterile neutrinos (3+2), and models in which the high mass of the sterile neutrino causes it to decay (3+1+decay.) The projected oscillation measurements can be see in \figref{wiggles}.

\section{Overview of Other BSM physics allowed by this design}

While IsoDAR is designed to be a neutrino source, the uniqueness of its design leads to access to other BSM physics processes. This is augmented by the excellent energy resolution and low backgrounds of the Yemilab detector.   BSM process can be explored deviations of the electron-neutrino elastic scattering cross section and through bump hunts of new particles that are produced in the IsoDAR target.

In addition to IBD events, the IsoDAR experiment is projected to observe $\sim$7000 $\bar \nu_e$-$e^-$ elastic scattering events over 3 MeV in a 4 year livetime. This sample can be used to search for non-standard neutrino interactions (NSIs) by testing the right and left handed couplings of  $g_L$ and $g_R$. These couplings are suspected to be modified by an effective coupling, $\epsilon$, which is dependent upon the flavor of the interacting particles. With our count of electron scattering events will be able to provide a precision measurement of $\epsilon_{eeL}$ and $\epsilon_{eeR}$, which would modify the Standard Model couplings by $g_R^\prime =g_R + \epsilon_{eeR}$ and $g_L^\prime =1 +g_L + \epsilon_{eeL}$. Obviously, if the Standard Model predictions are correct, $\epsilon$ = (0,0). However, any deviation from this would be indicative of new physics. While several programs using $\pi$ or $\mu$ decay-at-rest or decay-in-flight beams plan on also investigating this, such as groups at Fermilab or los Alamos, these investigate a different allowed region than the pure anti-electron neutrino beam of IsoDAR. Reactor experiments are able to probe the same channel, but have insufficient statistics to compete with IsoDAR@Yemilab. Sensitivities to  $\epsilon_{eeL}$ and $\epsilon_{eeR}$ as well as a more in depth explanation of this process can be seen in \cite{alonso2022neutrino, alonso2021isodar}.

Photons are produced in the IsoDAR target by various processes including nuclear excitations which provide several distinct photon peaks. In many theoretical models, photons couple to various particles such as X mass mediators and axions. 

X mass mediators can address two different anomalies- first the Atomki, or X17 anomaly in which there was observed a 6.8 $\sigma$ excess of lepton pairs at high opening angles in the decay of $^{8}$Be \cite{aleksejevs2021standard,krasznahorkay2019new}. To explain this anomaly, several groups have proposed a hypothetical boson with mass 17 MeV \cite{zhang2021can,delle2019atomki,zhang2021can}. Typically, experiments testing this result require detectors with high position resolution to observe the angular dependence of the decays. However, due to the nuclear transitions present in the IsoDAR target, as well as its close proximity to a kiloton scale detector, it is possible to investigate these bosons in a new way. Using IsoDAR it is possible to do a bump hunt to search for an energy signature of the particle at 17 MeV \cite{alonso2022neutrino}. Secondly, IsoDAR can be used to explore the 5 MeV bump, which has been previously described in this section. It has been proposed that the bump is due to miscalculations in the nuclear neutrino model. In this scenario, IsoDAR should not observe a 5 MeV bump because the lithium beta decay spectrum is well understood. However, the shape of the 5 MeV bump can be replicated at some masses of a potential X mediator boson. If this is observed, it gives strong evidence for an X mediator particle, if not, it helps identify the source of the 5 MeV reactor bump, greatly improving the physics case for IsoDAR.

\begin{figure}[!tb]
    \center{\includegraphics[width=\textwidth]
    {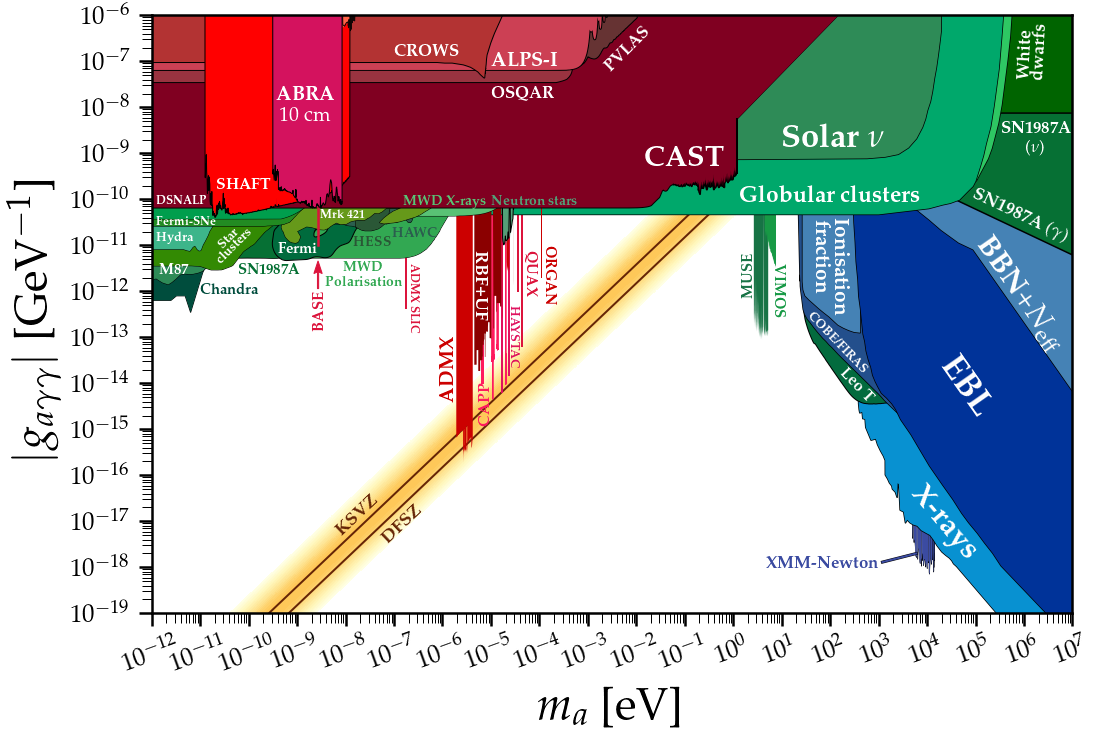}}
    \caption{\label{fig:axions_all} diagram of explored parameter space of axion comparing axion mass to coupling to the electromagnetic field. From \cite{AxionLimits}}
\end{figure}

The large flux of photons is also able to couple to the theoretical dark matter particle known as the axion. This particle was originally proposed as a solution to the fine-tuning Strong CP problem, but has since grown to be a compelling dark matter candidate. The axion can be described by its mass and its coupling constant, and based on these parameters may explain some or all of the currently observed dark matter. Many experiments take advantage of the axion-photon coupling. A plot describing this space can be seen in \figref{fig:axions_all}.

Due to the high background at low energies in the IsoDAR experiment, any axion searches in IsoDAR are restricted to be above a mass of 3 MeV. However, in the region known as the "cosmological triangle" there is still unexplored parameter space describing a QCD and dark matter axion.Using the photons produced in the IsoDAR target it is possible to confirm and compete with some of these experiments at higher masses. Due to the sharp nuclear peaks that provide clear signal above background, IsoDAR is competitive with beam dump experiments that have much higher power.  

More importantly, the nuclear transitions are also able to produce axions via the axion-nucleon coupling. This provides a much higher flux, and allows IsoDAR to probe new parameter space for nucleon couplings up to masses at 10 MeV. Similarly, there is a large electron and positron flux that results from the nuclear reactions that occur within the target. Using GEANT4 we simulated the photon and lepton flux within IsoDAR, and used this spectrum to calculate potential event rates for these theoretical particles.

A more detailed discussion of axions is available in the next section, and a discussion of X mass mediator particles can be seen in \cite{alonso2022neutrino}.

%% file: chap3.tex
\chapter{Searching for Axion-Like Particles using IsoDAR}

A primary result of this thesis is the calculation of sensitivity to axion production at IsoDAR.   This experiment joins many others across a wide range of coupling and mass parameters for the axion.    Before presenting the IsoDAR capability, we first very briefly review the many other approaches to axion searches.

\section{Modern Axion Experiments, in Brief}

To fully explore the possible properties of the axion, multiple experiments have searched the parameter space describing and surrounding the QCD axion. The can be searched for by several methods. These include cavities, such as  ADMX~\cite{Asztalos:2001tf,Du:2018uak}, and HAYSTAC \cite{Brubaker:2016ktl,Droster:2019fur}, LC resonators such as ABRACADABRA~\cite{Kahn:2016aff,Salemi:2019xgl}, DM radio~\cite{silva2016design} and SHAFT~\cite{gramolin2021search}, helioscopes including CAST~\cite{Zioutas:1998cc,Anastassopoulos:2017ftl}, and IAXO \cite{Irastorza:2013dav,IAXO:2019mpb}, light-shining-through-wall type experiments such as ALPS II \cite{Spector:2019ooq}, reactor experiments such as MINER, CONUS, TEXONO ~\cite{Dent:2019ueq,AristizabalSierra:2020rom,Chang:2006ug}), dark matter experiments such as XMASS \cite{Oka:2017rnn}, EDELWEISS \cite{Armengaud:2013rta,Armengaud:2018cuy},  SuperCDMS~\cite{PhysRevD.101.052008}, XENON~\cite{Aprile:2020tmw,Dent:2020jhf},  PandaX~\cite{Fu_2017}), searches for resonant absorption by nuclei ~\cite{Moriyama:1995bz,Krcmar:1998xn,Krcmar:2001si,Derbin:2009jw,Gavrilyuk:2018jdi,Creswick:2018stb,Li:2015tsa,Li:2015tyq,Benato:2018ijc,Dent:2021jnf}, constraints made by astrophysical observations ~\cite{Dent:2020qev, Carenza:2021alz, Galanti:2018nvl,Tavecchio:2012um,Galanti:2015rda,Ayala:2014pea,Fermi-LAT:2016nkz,Conlon:2013txa,Conlon:2015uwa,Conlon:2017qcw,Raffelt:1987im}, and accelerator based experiments such as  NA62 \cite{Volpe:2019nzt}, NA64~\cite{Dusaev:2020gxi,Banerjee:2020fue}, FASER \cite{Feng:2018noy}, LDMX \cite{Berlin:2018bsc,Akesson:2018vlm},  SeaQuest~\cite{Berlin:2018pwi}, SHiP \cite{Alekhin:2015byh}, PASSAT~\cite{Bonivento:2019sri}).

\subsubsection{Cavity Detectors}

\begin{figure}[tb]
    \center{\includegraphics[width=.7\textwidth]
    {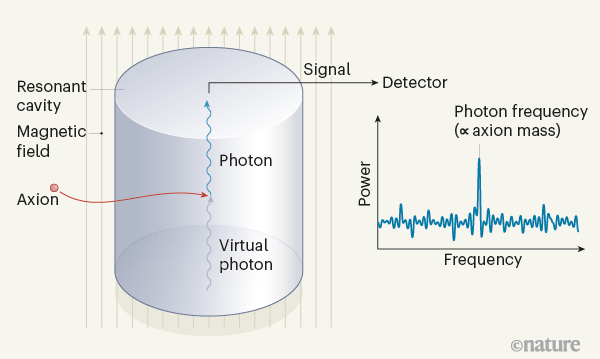}}
    \caption{\label{fig:haloscope} Cartoon showing the function of a haloscope axion detector. Ambient axions travel through the microwave resonant cavity which is held within a magnetic field. The axion scatters producing a detectable photon. When the cavity is properly tuned, the photon is observed as a resonant signal in the microwave cavity, which is proportional to the axion mass. Taken from \cite{backes2021quantum}}
\end{figure}

Cavity detectors search for ambient axions whose path goes through a detector. These experiments take advantage of the earth traveling through the dark matter halo of the milky way. Cavity detectors use powerful magnetic fields to convert axions into detectable microwave photons. Typically a haloscope experiment is made up of a large magnet, microwave cavity, and quantum electronics. These experiments expect low event rates, and thus require extremely low backgrounds and noise. This often requires ultra-cold setups and electronics to reduce thermal noise.

The microwave cavity within the detector can be tuned to different frequencies corresponding to different axion masses. These experiments therefore search for axions in "slices" of parameter space. In the case of the haloscope, different values of $g_{a\gamma}$  influence the strength of the signal, not whether the signal is present. Therefore in order to reach lower couplings, the experiments are typically limited by signal to noise ratio, not by axion flux. These experiments can then tuned over a range in order to find an axion signal. This can be seen in  \figref{fig:haloscope}

\subsubsection{LC resonators}
ABRACADABRA ~\cite{Kahn:2016aff,Salemi:2019xgl} is an LC resonator that investigates sub- $\mu$-eV scale axion masses using a SQUID magnetometer. ABRACADABRA uses a toroidal magnet to produce the needed magnetic field, and a superconducting pickup circuit in its center that precisely measures the magnetic flux.

While cavity experiments such as ADMX \cite{braine2020extended} can take advantage of the boundary conditions of the cavity, this limits the investigated masses to shorter wavelengths. 
The advantage of the setup of ABRACADABRA and DM radio is that without the use of a cavity they are constrained to specific wavelengths, and thus are able to probe wavelengths that are much longer.

\subsubsection{Helioscopes}
Helioscopes operate under a similar concept to Haloscopes, except that they specifically observe axions produced in the sun. Solar axions are expected to be produced via Compton scatters, i.e. $e + \gamma \rightarrow e + a$ and axion bremsstrahlung i.e. $e + Z \rightarrow e + a +Z$. The flux can then be modelled based on solar axion models \cite{hoof2021quantifying,schlattl1999helioseismological,gao2020reexamining}.

\begin{figure}[tbp]
    \center{\includegraphics[width=.6\textwidth]
    {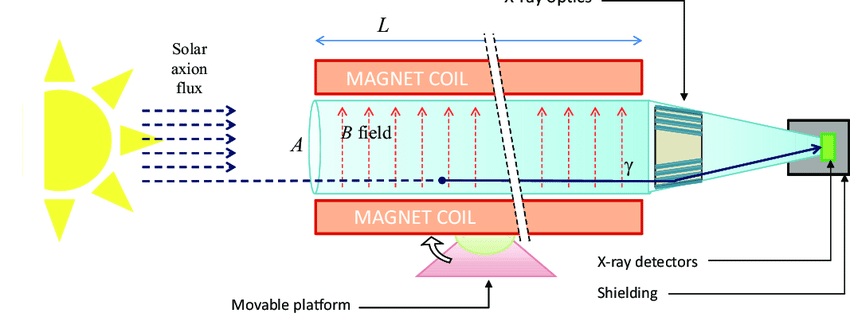}}
    \caption{\label{fig:helioscope} Cartoon showing the layout of a helioscope axion detector. Axions produced in the sun are converted to detectable x-rays inside a large magnetic field. The helioscope is pointed away from the sun via its mobile platform to observe backgrounds. Taken from \cite{vogel2015next}}
\end{figure}

The advantage of a helioscope being that there is less concern over signal to noise than a haloscope. These experiments are also typically much larger. Helioscopes use magnetic fields to produce photons via axion scattering, but rather than using a microwave cavity within the magnetic field, there is an x-ray detector placed at the end of a magnetic field path, as seen in \figref{fig:helioscope}.  In the case of CAST \cite{Zioutas:1998cc,Anastassopoulos:2017ftl}, a 9 T decommissioned LHC test magnet is used.

The uniqueness of a helioscope is the expected difference in signal whether the detector is facing towards or away from the sun, creating a clear signal compared to the backgrounds which are taken at night. Therefore, it is possible to isolate events in the detector as those influenced by the sun. This allows exploration of a wide range of axion masses.

\subsubsection{Light Through Walls}

\begin{figure}[tbp]
    \center{\includegraphics[width=.6\textwidth]
    {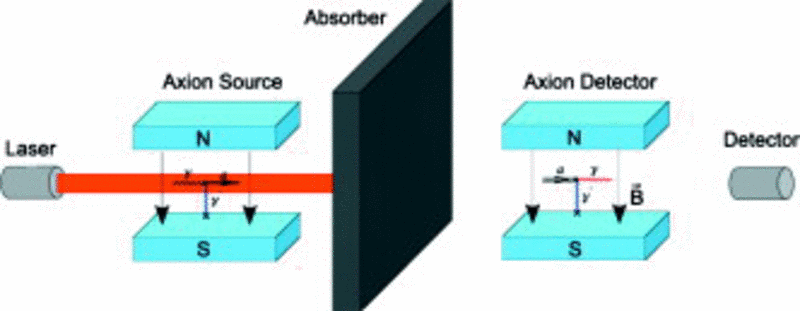}}
    \caption{\label{fig:light_through_walls} Cartoon showing the layout of a light shining through walls type experiment. Some form of light source is impidgent on an absorber after going through a magnetic field. When in the magnetic field, the photon may scatter to produce axions, which will travel through the absorber but the photons will not. When traveling through the magnetic field on the other side, the axions may produce photons, which may then be detected. Taken from \cite{kim2010axions}}
\end{figure}

A common paradigm of axion experiments is the "light through walls" experiment. This is when a high power light source is placed on one side of an absorber and a sensitive photodetector is placed on the other. While light is absorbed by the material, axions would be unlikely to interact and thus traverse the wall. Axions may then convert back into detectable photons in the magnetic field on the opposite side of the wall. These photons can then be observed up by a photodetector. The light source used is typically thought of as a laser, however, nuclear reactors or accelerators can also produce a high flux of photons at much higher energies than most lasers. This makes them an attractive alternative. Similarly, the absorber can be nuclear shielding or any obstructing material that does not allow photon transmission. This makes this experiment an easily generalizable example. A diagram can be seen in \figref{fig:light_through_walls}.

\subsubsection{Nuclear Reactor Based Experiments}
The nuclear reactions that occur to produce power within a nuclear reactor produce a high flux of photons as well. These photons are unable to pass through the walls and shielding that surround the nuclear reactor, however it is possible for the photons  to Primakoff scatter to produce axions.

These axions could then Primakoff scatter to produce photons within a detector. A sufficiently compact detector can be put relatively close to the reactors to investigate this possibility. These detectors are typically used for neutrino measurement experiments at reactors, however due to the similar signature that occurs with Primakoff scattering many of these detectors may be used for axion experiments as well.

The advantage of using nuclear reactors is that their production of a high flux of axions at a relatively low cost while the reactor is running. However, the disadvantage is that these reactors fluxes are not always well understood due to their industrial use. The detectors which are used at these reactors also must be quite small, as they must be brought to the reactor and not interfere with daily operations. The photon spectrum, while having a high flux, also has a sharp cutoff typically sharp.  \cite{Dent:2019ueq,AristizabalSierra:2020rom,Chang:2006ug,agnolet2017background,brdar2021axionlike}. So while able to exclude certain areas at high confidence, reactor type experiments are not necessarily able to explore a large parameter space for the axion. These experiments are very well suited within certain regions, such MeV scale axions. In some cases, these experiments can also be used to study axion-nucleon couplings, which will be explained later in this chapter.

\subsubsection{Dark Matter Experiments}
Experiments searching for WIMP dark matter often have large liquid scintillator detectors in order to detect recoils and excitations from WIMP interactions \cite{Aprile:2020tmw,Dent:2020jhf}. However, these experiments may also be used for axion-like particle searches with low backgrounds. 

These large dark matter detectors make use of the axio-electric effect \cite{derevianko2010axio} and the cross section for producing photons. Axions that interact in this way produce high energy gamma rays that shower and provide a detectable signal.

\subsubsection{Resonant Absorption by Nuclei}
It is possible to put limits on solar axions by observing the photons that result from the de-excitations of nuclei. If solar axions are present, there would be a difference in the number de-excitation photons from the nucleus as axions are absorbed and excite the atom. 
These experiments not only probe the coupling to photons, $g_{a\gamma}$ but also the axion coupling to nucleons, $g_{aN}$, and the coupling to electrons $g_{ae}$. Some experiments using primakoff production mechanisms for solar axions\cite{Derbin:2009jw, di2022probing}, others are able to circumvent the use of the electromagnetic coupling entirely \cite{Moriyama:1995bz}.

While often used to describe solar axion experiments, these same couplings can be observed in other sources of nuclear reactions, such as reactors, accelerators or specialized experiments. \cite{aguilar2022first,dent2020new, lee2019new, aguilar2021axion}.  Taking advantage of these nuclear resonances provide an important new avenue to search for axion like particles.

\subsubsection{Constraints from Astrophysical Observations}

Methods of astrophysical limits are highly varied based on the region that is being investigate and the model being used. For the sake of this work I will focus the region surrounded by the cosmological triangle in the MeV mass range, which is limited by:
\begin{itemize}
    \item The lifetime of horizontal branch stars
    \item Supernova 1987a
\end{itemize}

The existence of axions would have a significant impact on the evolution of horizontal branch stars. The existence of axions could reduce the lifetime of these stars as they transfer energy away at an observable rate, while only negligibly changing RGB evolution \cite{raffelt1990astrophysical, salaris2004initial}. However, this is not the case, as current models are able to reproduce R values in stellar evolution within 30\%, excluding a large amount of axion parameter space.

One issue of using the limits set by horizontal branch stars is the high complexity and variability of the stellar evolution models. Because these models describe massive ensembles of highly complex nuclear reaction chains, they do not have the advantage of a well controlled system such as that in a lab experiment. 

Supernova 1987a describes an observation in which a massive progenitor star underwent a core collapse, then explosion. The shock wave expelled the inner parts of the star, including its mantle and envelope. Neutrinos escaped the gravitational pull, and release a large amount of the energy in the first few seconds. (This is why neutrino detectors are often used as early warning systems for supernova observations.)

Based on neutrino measurements we can estimate the luminosity of supernova 1987a, and thus the axion flux at earth\cite{lee2018revisiting}. In order to properly evaluate the multiple axion models in this case, the couplings to nucleons and photons must be evaluated independently. 

The arrival time of an axion in the detector is dependent on its mass. Because the neutrino is nearly massless, at sufficiently high mass any signal from an axion would be delayed from the earlier neutrino signal from the supernova. Because this signal was not observed, a large mass range of axions can be eliminated as long as the coupling is sufficiently high for the axion signal to be detected.

\subsubsection{Accelerator Based Experiments}

High energy photons are produced in nuclear reactions that can be initiated by proton beams on target. The advantage of this method is that the proton beam can be produced at much higher energies, producing a high flux of photons. This has been done at SLAC, CERN, as well as several accelerator based experiments\cite{dobrich2017axion,lu1987search, riordan1987search, hook2020high,ren2021detecting, balkin2022probing,Volpe:2019nzt,Dusaev:2020gxi,Banerjee:2020fue,Feng:2018noy,Berlin:2018bsc,Akesson:2018vlm,Berlin:2018pwi,Alekhin:2015byh,Bonivento:2019sri}. Typically, these experiments probe higher masses in the MeV- GeV region. An advantage of using proton beams on target is that they can lead to nuclear resonances, which as previously stated can be used as a means for axion production. Simultaneously, this resonances provide sharp photon peaks which provide a clear signal above background for axion production. Lastly, these violent interactions release large numbers of electrons and positrons, which can then be used to observe axions via their electron couplings. The disadvantage is that these linear accelerator systems are very expensive, large, and typically focused on another primary experiment. At higher energies some of the signal can also be washed out by hadronic and electromagnetic cascades. This prevents these experiments from being coupled to many detectors that would have low backgrounds needed for axion experiments, for instance an underground detector.

A compact accelerator such as a cyclotron can be used to produce a proton beam in close proximity to an underground detector in a relatively inexpensive manner.

 The disadvantage is that this method produces much lower fluxes than high power synchrotrons or linacs.
However, coupling a cyclotron with an underground detector (such as the LSC at Yemialb) allows for very low backgrounds and high sensitivity, since there are negligible other particles in between the detector and the beamline. Backgrounds from low energy cosmic rays are filtered out by  shielding around the detector. 
Building these detectors is highly expensive, which is why it is common to use these detectors for multiple experiments simultaneously.This allows a compact experiment to run with higher statistics than would otherwise be able to be done due to budget constraints.
The optimal solution thus would be a compact accelerator system that is able to produce a high photon flux near an underground detector. This could potentially explore new parameter spaces that have not been able to be excluded by other similar experiments.

To approach this ideal system, we propose the IsoDAR experiment at Yemilab as a search for axion-like particles. While this is still a parasitic measurement on a primarily neutrino experiment, in the following publication we show that this compact, inexpensive experiment is capable of exploring new parameter space as well as duplicating several non-lab based limits. This is done by taking advantage of the nuclear resonances that occur in target during the proton bombardment from the IsoDAR cyclotron.

\section{IsoDAR and the search for Axion-like Particles}
\includepdf[pages={-}]{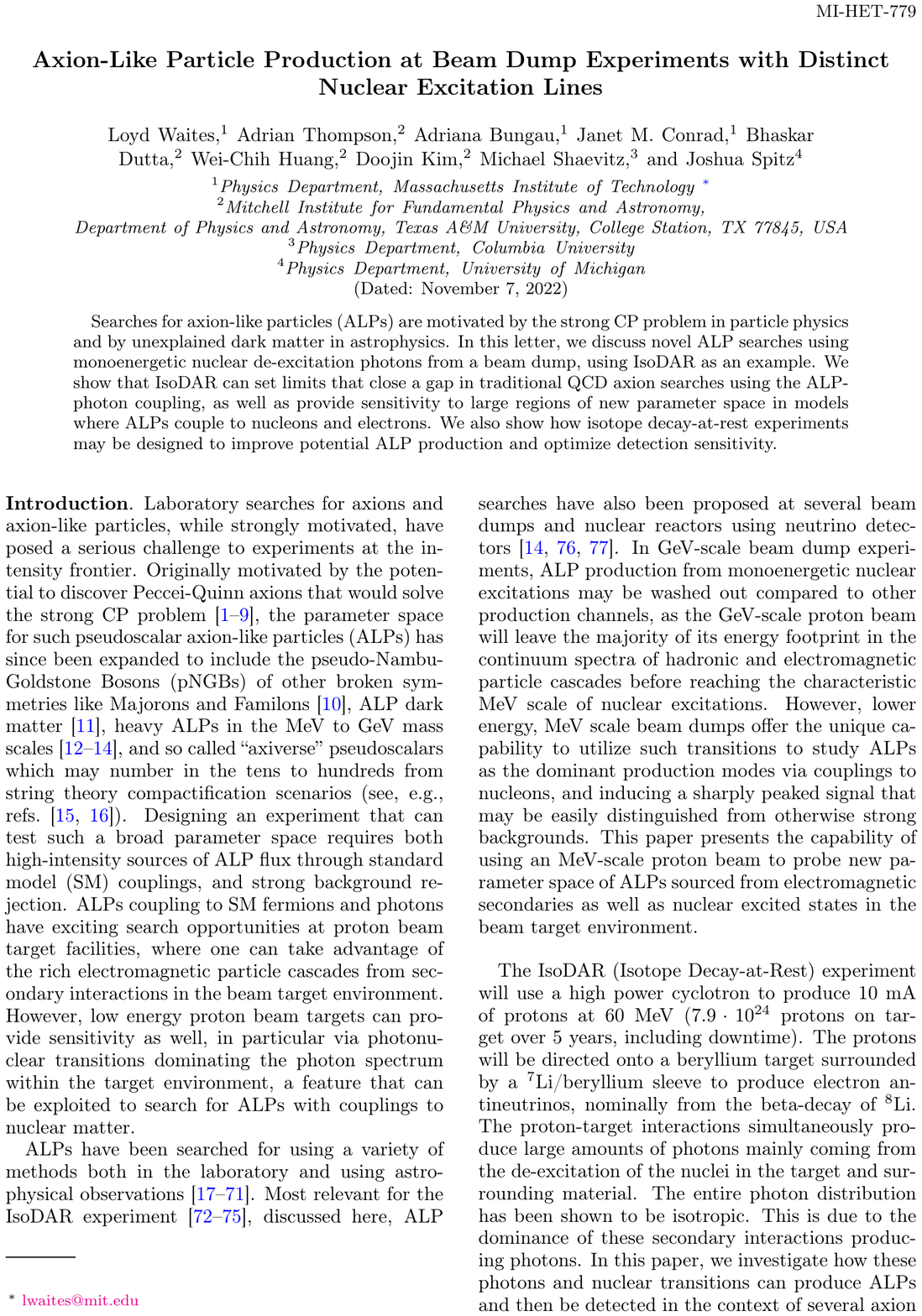}

%% file: chap4.tex
\chapter{The Physics of the IsoDAR Accelerating System}

\section{Introduction to Accelerator Physics}

\subsection{Space Charge}

When ions come in close proximity to each other, they are repelled electromagnetically by Coulomb repulsion. Therefore, a beam of ions grows over time as the ions push each other apart electrically and change the quality of the beam. This behavior is called space charge.

Space charge can have an important impact on the shape of a beam. Space charge forces expand the beam radially as well as longitudinally. When the bunch-charge-distribution is not uniform, this force is non-linear force thus causes emittance growth. In order to overcome the expanding forces of space charge, electromagnetic focusing forces are required to keep the beam in the desired shape. 

Ion beams are almost exclusively transported in vacuum in order to prevent interactions with the air; however, residual gas remains within the beamline. The residual gas is ionized and secondary electrons are mixed into the beam. When the ion beam is positively charged, attractive forces from the electrons counteract the Coulomb repulsion within the current of the beam, lessening the effects of space charge. This process is thus called space charge compensation. While we do understand this physically, it is difficult to predict due to the need for a precise knowledge of the vacuum of one's system, and so it is most often measured in experiment.

Intuitively at higher currents space charge forces increase, decreasing beam quality and turn separation. To overcome these forces, we will use new methods to achieve the high currents needed for IsoDAR.

\subsection{Phase Acceptance}

 When designing an accelerator, it is common to think of a "test particle," or a single particle that follows the ideal path through the accelerator with ideal energy and phase. However, a beam is an ensemble of particles. The ensemble of a beam at a given slice can be described by each particle's energy and phase.

Acceleration within a cyclotron can only occur when the bunching frequency is precisely matched by the cyclotron frequency. A boundary is formed in this energy-phase diagram in which particles are no longer accelerated, and are lost. Particles outside this boundary are "out of sync" with the RF phase of the cyclotron. This boundary is known as a separatrix. 

When the ensemble is made close to the test particle, it has the flexibility to have error in its position in phase space while still remaining within the separatrix, continuing to be accelerated. This is known as phase stability. 

To maximize the transmission through a cyclotron, it is therefore advantageous to synchronously bunch a beam before its injection. Each of these bunches is able to be within this phase acceptance window, rather than with DC beam injection in which the uniform phase of the beam leads to large amounts of beam existing outside of the separatrix. 

To maximize power from a cyclotron, it is not only important to maximize transmission (and thus the amount of beam within a phase window), but to fill each RF bucket. This occurs when the frequency of the bunches is matched to the frequency of the cyclotron, so the phase window isn't only filled, but EVERY phase window is filled. This is know as continuous-wave or CW operation. 

As a side note- it is also possible to lower the power of a cyclotron by only filling not being in CW operation but remaining in phase, i.e. filling every other RF bucket. This can be used for early stages of commissioning.

\subsection{Vortex Motion}
\begin{figure}[tpb]
        \center{\includegraphics[width=.8\textwidth]
         {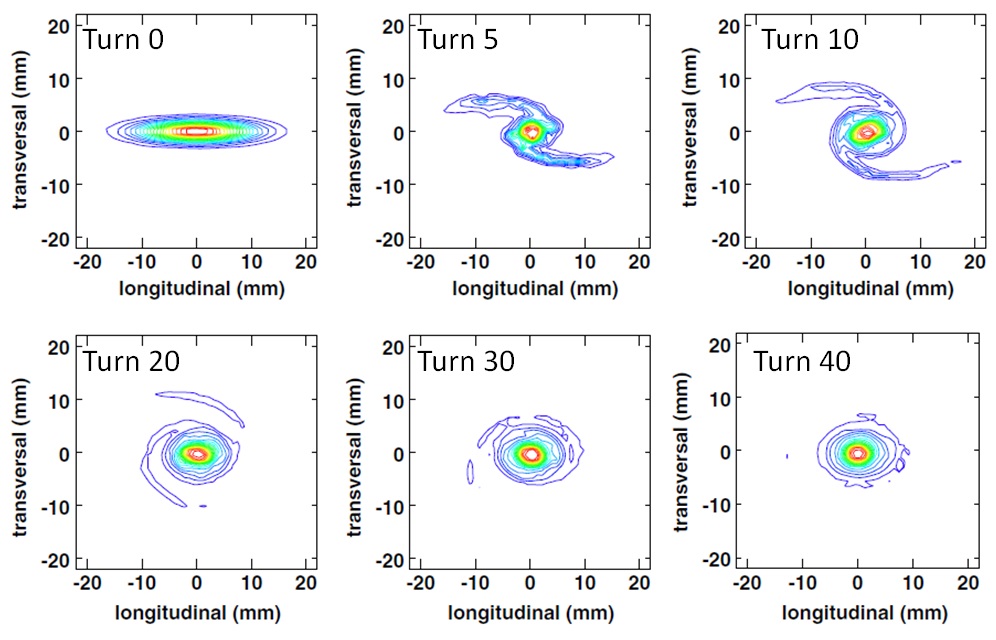}
         
        \caption{\label{fig:vortex_psi} Vortex motion simulations done at PSI show the evolution of the bunch from the injection into the cyclotron to the 40th turn. Taken from \cite{yang2010beam}}}
      \end{figure}

It was found at the PSI Injector II cyclotron that at high currents space charge causes nonlinear effects that lead to a round shape in the beam profile \cite{yang2010beam}. This can be seen in Figure \ref{fig:vortex_psi}

As mentioned previously, space charge causes a radial force on the beam that grows with current. The magnetic field from the cyclotron is along the vertical axis, and the bunch is moving in a spiral around it. Each particle within the bunch is spreading out relative to the center of the bunch due to Coulomb repulsion. The Lorentz force acts in the direction of the cross product of the velocity of the particle and the magnetic field. This causes the particles to rotate within the bunch, redirecting the radial force and keeping the bunch together. This keeps the bunch round, rather than going through a type of propellering motion. This allows for cleaner extraction and higher turn separation. Because of this intrinsic rotation, this effect is called "vortex motion."
This effect is only significant at reasonably high currents. Vortex motion was described in depth by PSI \cite{yang:cyclotron_sim, stetson:strong_bunching, stammbach:strong_bunching,baumgarten:vortex}.

\section{The challenges of designing for 10~mA at 60~MeV}

The requirements of the IsoDAR accelerator in order to complete the experiments mentioned in Section 2 and 3 above are as follows:
\begin{enumerate}
    \item An accelerator system that can produce 10~mA cw of 60~MeV protons on target.
    \item The accelerator system must be sufficiently compact to fit in an underground facility with minimal excavation, and be designed in such a way that it may be assembled in such a location.
    \item Accelerator system must be constructed at a reasonable cost.

\end{enumerate}

In order to address points 2 and 3, the IsoDAR collaboration decided to use a compact cyclotron. Cyclotrons provide compact systems that are typically less expensive than linear accelerator systems. 

Addressing point 1 for a cyclotron system is a far greater challenge. The record holder for highest current cyclotron is the PSI Injector II cyclotron at $\sim$3mA \cite{kolano2013precise, stetson1992commissioning,stammbach2001psi,yang2013beam}. However, not only is this 3~mA insufficient for our experiment, the PSI Injector II cyclotron is also a separated sector cyclotron. These types of cyclotrons are far larger and more expensive than the typical commercial cyclotrons.

To ensure a more compact system, commercial cyclotrons are not separated sector machines, however their currents are on order 1~mA \cite{waites2020potential}. These systems are far too low in power to achieve the goals of IsoDAR within a reasonable timescale.

We needed to design a new cyclotron to meet these criteria. In order to make a high current, compact cyclotron we take advantage of several new accelerator techniques.

\section{Path to higher currents}
To develop a high power, compact cyclotron, three developments were used in cyclotron technology:
\begin{enumerate}
    \item Use of \htp to mitigate space charge.
  \item Using an RFQ direct injection system to have high transmission and limit space charge effects in low energy regions. 
    \item Designing to take advantage of the vortex motion effect in order to provide a clean extraction.

\end{enumerate}

The spiral inflector is the point in the central region in which the beam is injected, and changes planes from being axial to the cyclotron, to being in the midplane. This is done using two spiral electrodes held at a high potential difference, creating an electric force which manipulates the beam into the desired orientation.
In order to properly simulate this region, the particle in cell (PIC) code OPAL \cite{adelmann2008opal, winklehner2017realistic} was used. OPAL was developed specifically for accelerator applications and is highly parallelizable. 
This code has been benchmarked against experimental results and has shown to accurately reproduce the trajectories of particles through the spiral inflector.

During extraction from the cyclotron, the beam  enters an electrostatic extraction channel comprising a septum and a negative puller electrode, which further increases turn separation by applying a radial electric field on the order of 2 MV/m. 
A concern for this septum is that it will be radio-activated upon having high currents impinge upon it. It is thus crucial to ensure highest possible turn separation on the final turns of our cyclotron, as any overlap would collide with the septum.

In an H$^{-}$ cyclotron, it is possible to compensate for poor turn separation in using a stripping extraction over a larger area \cite{grillenberger2021high, baartman1995intensity,schneider1988compact,baartman2015space}. However, this process results in a large energy spread in the beam. This process also does not provide the limited space charge benefits of \htp. In addition to making the beam more difficult to control, the beam will break into small beamlets that each exhibit vortex motion, rather than a single, clean vortex.

\begin{figure}[pbt]
    \center{\includegraphics[width=.8\textwidth]
    {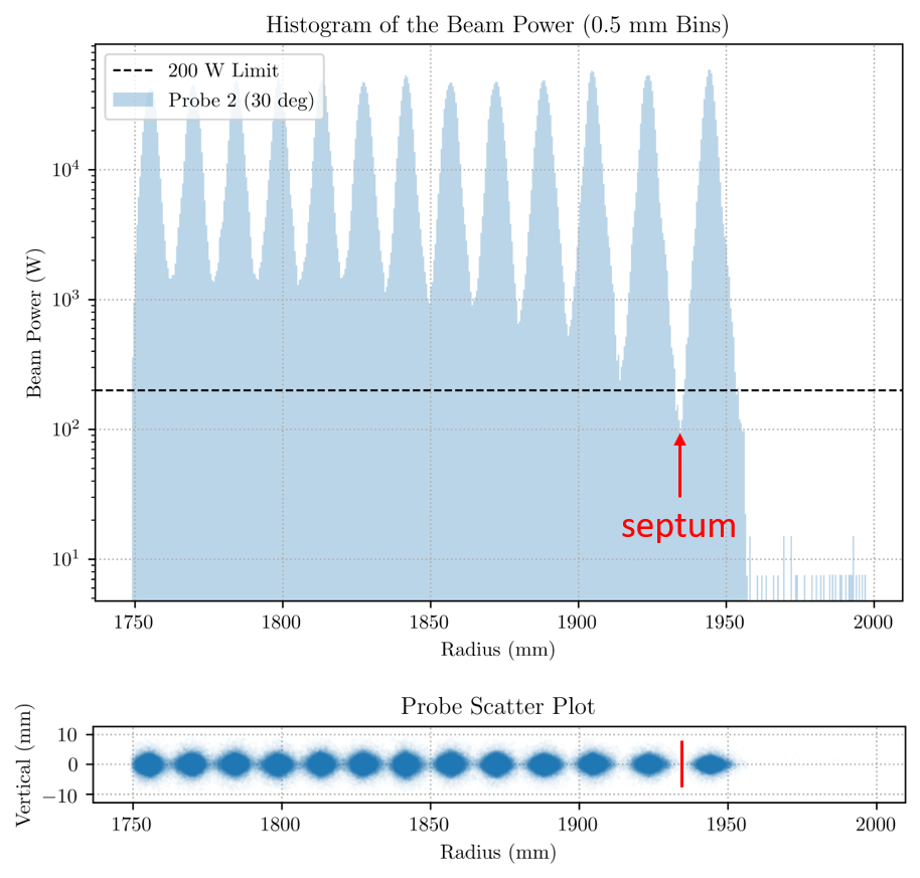}}
    \caption{\label{fig:Power_septum} Simulations describing the turn separation of the outer orbits in the IsoDAR cyclotron. This shows that power below acceptable levels will be deposited on the septum. Note the logarithmic scale. Taken from \cite{alonso2021isodar}}
\end{figure}

\section{Use of \htp}

Typical commercial proton cyclotrons accelerate H$^{-}$ ions, however; these cyclotrons are unable to reach currents higher than about 1~mA due to issues with space charge\cite{alonso:isotopes, baartman1995intensity}. An IsoDAR collaborator from INFN-Catania, Luciano Calabretta, found that by replacing the H$^{-}$ with H$^{+}_{2}$ that the accelerated current would be a factor of two less than an analogous proton current, reducing space charge while in the cyclotron \cite{calabretta:h2+}. After the extraction, the H$^{+}_{2}$ could be stripped of electrons, and the current would be doubled as the H$^{+}_{2}$  was broken up into protons.

Our cyclotron also requires the use low-emittance, high power, and highly pure \htp ion source. This ion source is crucial in order to maintain the needed currents and quality for our beam. This ion source and associated system will be described in more detail in the following chapter and appendix A.

The use of \htp comes with an additional benefit. While simulations show the power on septum is at acceptable limits, as see in Fig. \ref{fig:Power_septum}, we also have the option to place a stripping foil to protect the septum. The stripping foil removes an electron from the \htp, breaking the beam into protons. With the protons having a different charge to mass ratio than the \htp, the protons then follow a different trajectory, and do not collide with the septum. These protons can be extracted separately, and used for an additional project such as medical isotope production \cite{waites2020potential}. This is made possible by the use of \htp in the IsoDAR cyclotron.

\section{RFQ direct Injection}
While we use an RFQ for slight acceleration of our beam into the cyclotron, the primary advantage of RFQ direct injection is that it allows for significantly higher transmission by bunching the beam, allowing for a higher amount of beam to be in the desired phase acceptance window.

The IsoDAR cyclotron requires a bunched beam at 32.8 MHz \cite{alonso2021isodar}. These bunches are then accelerated by "riding the wave" of the alternating RF voltage. However, the ion source creates a DC beam of ions. If this was directly injected into the cyclotron, the beam would only have approximately 5\% transmission efficiency because the RF system would only accelerate +/- 10 degrees of the synchronous RF phase. If there is an RF buncher added, which is common practice, the transmission  efficiency can be increased to up to 15\%.  

Using a radio-frequency quadrupole (RFQ) we can take the injected beam from the ion source, and tune it into bunches with 32.8 MHz frequency in order to match with the cyclotron frequency. By using an RFQ to heavily bunch the beam this way we expect to have up to a 99\% transmission efficiency from our ion source to the central region of the cyclotron. More details on the RFQ will be explored in Chapter 6.

\section{Designing with Vortex Motion}

In the IsoDAR cyclotron, vortex motion has also been studied extensively. We used the above mentioned OPAL codes that were also used at PSI for the Injector II simulations for our own simulations. OPAL was extensively benchmarked against other codes and experiments at PSI Injector II, giving us a high degree of confidence in the IsoDAR simulation results~\cite{winklehner2021order}. We have shown an example of these simulations in \ref{fig:vortex_IsoDAR}. Many of the details of the full IsoDAR design and simulations can be found in appendix B.

 \begin{figure}[t]
        \center{\includegraphics[width=\textwidth]
         {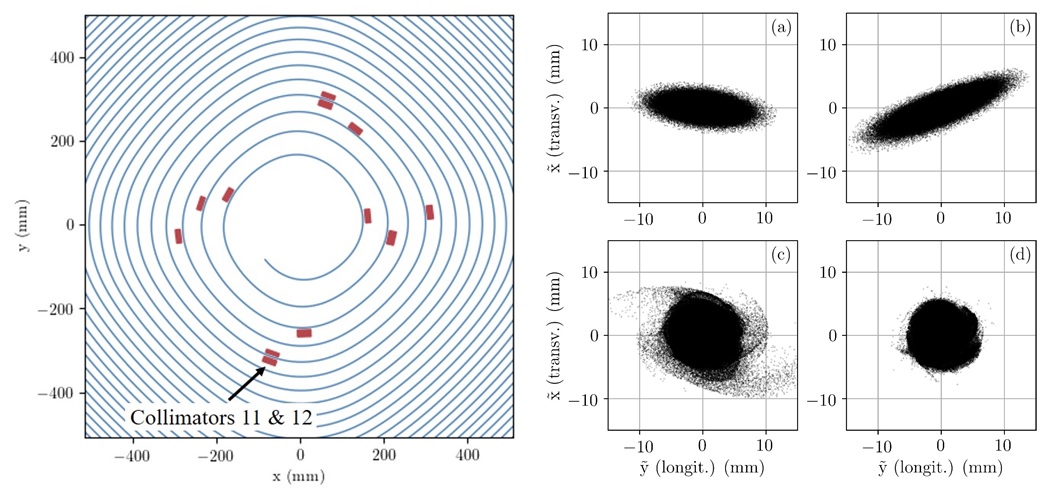}
         
        \caption{\label{fig:vortex_IsoDAR} (Left) Placement of 12 collimators throughout the central region of the IsoDAR cyclotron. (Right) Simulated beam cross section at several points in the cyclotron, showing the vortex motion effect. Taken from \cite{winklehner2022order} }}
      \end{figure}

%% file: chap5.tex
\chapter{Design and Performance of the Prototype Ion Source}

As has been described, ion beams are crucial to the particle physics community. While not often in the limelight, it is obvious that beam requires a source of ions. These ion sources are a subfield all their own.

Typically to produce these ions a chamber is pumped with gas and ionized with microwave radiation, heated filament, or other ionizing agent. In some cases the particles are molecular and broken down into single atoms to form an ion gas, or in other cases (such as in IsoDAR) the gas molecule keeps some electrons and remains intact, but ionized.

There are multiple different forms of ion sources, however the most common are a electron cyclotron resonance (ECR) ion source, and a filament driven ion source. 
An ECR ion source has a magnetic field surrounding a chamber, and uses microwave radiation driven at the resonant frequency of the electrons to strip them from the gas. These stripped electrons then collide with other atoms or molecules in the gas, leading to further ionization. The typical free parameters in an ECR ion source are the microwave power, the pressure of the gas used inside the source, and the magnetic field used to confine the plasma.

A filament driven ion source uses a wire of a material with a high melting point, such as tungsten, and drives a high current through it. The filament is held at a different potential than the chamber, leading the filament to emit electrons with sufficient energy to ionize the gas around it. The plasma is held in place by a magnetic field from either permanent magnets or a solenoid. The typical free parameters of a filament driven ion source include the confining magnetic field,  the pressure of the gas used inside the source, the heating current running through the filament, and the potential difference between the filament in the chamber, or the filament discharge voltage.

In general, filament driven ion sources are capable of lower emittances, while ECR ion sources are capable of higher total currents. \cite{winklehner2021high,winklehner:rfq2,winklehner:mist1,axani:mist1}

Ion sources can be separate or internal to the accelerator. For example, a cyclotron can have an ion source in its center directly attached to the beamline. This provides better vacuum sealing, compactness, and a closed, and tuned system which is very difficult to alter. An ion source can also be external, in which case the ion source exists on its own outside of the accelerator. The ions are then transported to the nearby accelerator via LEBT (or in the case of IsoDAR, via RFQ). The advantages of an external ion source is its modularity. An external ion source can be quickly and easily disassembled for repair or replacement. This is critical in a filament driven ion source because of when the filament burns out or becomes damaged after long term use. For accelerators that intend to be used consistently over a long period of time such as IsoDAR, filaments will be replaced regularly. 

Ion sources use a system to extract the ions from the source and into the beamline. This is often done with electrostatics, but can also be down with small magnets or solenoids. This begins the shaping of the beam as well as provides the first steps of acceleration. The extraction system then must be modified and tuned in order to match the desired parameters for the first step of the beam line. In most cases, this extraction system generates a DC beam of ions. This is unmatched to the bunched beam required by several accelerators. This is why in many cases, an additional buncher is required between extraction and injection into the primary accelerator.

The ions that have been extracted are not in a perfect vacuum (or else the ion beam could not exist!). There are free gas molecules that will also be ionized in the source. These impurities (often including nitrogen or oxygen from the air, or unwanted levels of ionization from the initially pumped gas) will often require separation of the beam to eliminate any unwanted species. This can be done by using a dipole magnet, which will separate the species based on gyromagnetic ratio, or by tools such as an RFQ. 

\section{Physics of \htp xIon Sources}
\begin{figure}[t]
    \center{\includegraphics[width=.5\textwidth]
    {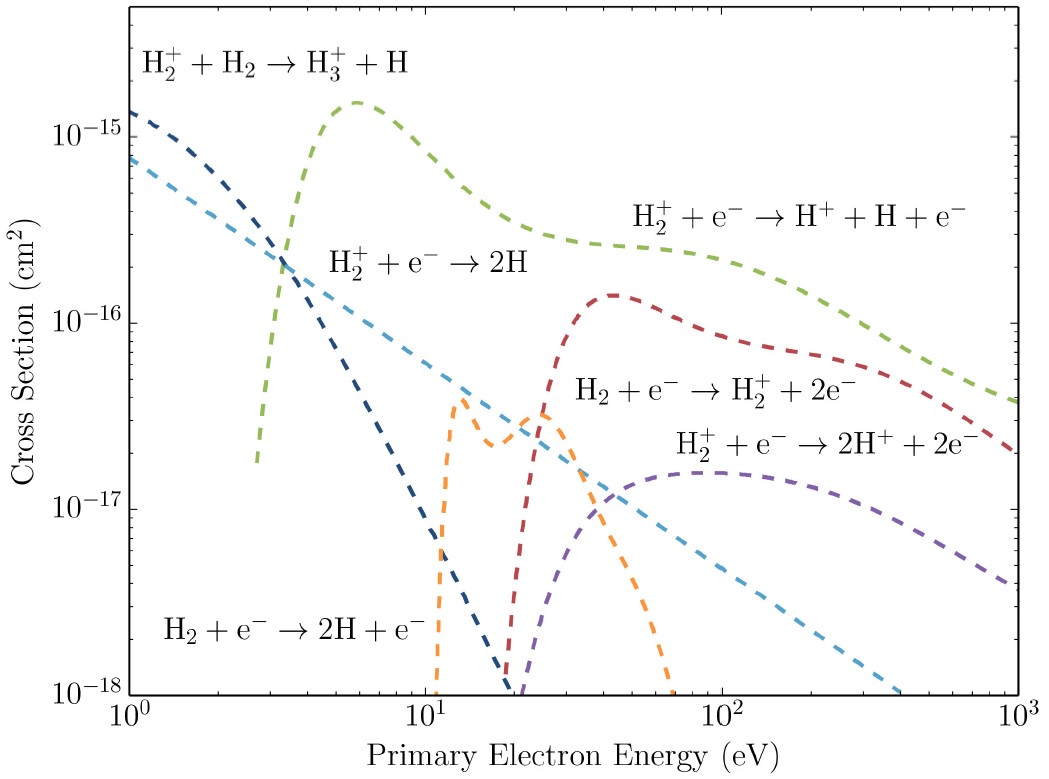}}
    \caption{\label{fig:htp_diss} Cross-sections of  
             production, recombination, and dissociation of \htp to other ions.
             From \cite{axani:mist2}}
\end{figure}

There are several competing processes in an ion source which uses hydrogen gas. Hydrogen ions are produced from electron-impact-ionization from  electrons which originate at the tungsten filament. The ions that are produced are determined by the electron energy (i.e. discharge voltage) and the cross section, which is a function of plasma density. The plasma density can be manipulated by varying the pressure inside the source with the MFC, or increasing the magnetic field to more tightly confine the plasma. \htp is primarily produced through the process:
\begin{equation}
    \hyd + \mathrm{e}^- \rightarrow \htp + 2 \mathrm{e}^-
\end{equation}
However, there are competing processes:
\BEA
    \htp + \mathrm{e}^- & \rightarrow & \hp + \mathrm{H} + \mathrm{e}^-, \\
    \htp + \hyd & \rightarrow & \hthp + \mathrm{H}, \\
    \htp + \mathrm{e}^- & \rightarrow & 2 \hp + 2 \mathrm{e}^-.
\EEA
So in order to tune an ion source to \htp is a delicate balancing act of ion source parameters. This is shown in more detail in \figref{fig:htp_diss}

\section{Requirements of Ion Source for IsoDAR}

In order to produce these high neutrino statistics proposed for IsoDAR, a high-power proton accelerator is required. To produce a high-power proton driver, it is necessary to have an ion source which is able to produce a high enough current to compensate for beam losses throughout the system. For IsoDAR we have chosen a filament driven multicusp ion source \cite{winklehner:nima} and so it is desirable to run at lower currents to limit the wear on the filament. The higher the wear on the filament, the more often the filament will require replacement. Higher currents in the system will also cause higher amounts of space charge, increasing emittance throughout the beam-line. This is particularly important in the low energy regions.  Therefore, making the transmission through the system as high as possible is crucial.

This type of ion source was chosen for its low emittance and high species fraction of \htp. \htp was chosen as the accelerated beam species for several reasons which were described in the previous chapter.

Contaminant ions in the beam can be filtered out by the RFQ due to their different charge to mass ratios, however over time this will cause damage to the RFQ electrodes. Therefore, it is important to minimize ion source contaminant species to prevent wear on the RFQ.

The ion source requirements for RFQ direct axial injection into the 
compact cyclotron are:
\begin{enumerate}
\setlength\itemsep{0.1em}
    \item Low emittance ($<0.1$ $\pi$-mm-mrad, rms, norm.)
    \item Low contamination ($>80$\,\% \htp fraction)
    \item High current (10$\sim$mA of \htp) 
\end{enumerate}
These stem from the desire to keep the system compact, and utilize the RFQ's ability
to separate by mass in addition to its highly efficient bunching and pre-acceleration.

\section{Design of MIST-1 ion source}

Due to the high current requirements of \htp required by IsoDAR, a new ion source has been constructed at MIT which is based on the design by Ehlers and Leung \cite{ehlers:multicusp1}. This ion source uses a tungsten filament to produce electrons which are accelerated by the discharge voltage ($\sim$100V). The collisions between these free electrons and the hydrogen molecules in the chamber ionize the hydrogen. The ionized hydrogen is confined using a multicusp magnetic field from a ring of permanent magnets that surround the chamber. To ensure that the ratio of \htp to \hp produced is favorable, the extraction region  has been made very short so that dissociation has insufficient time to occur. The source created by Ehlers and Leung had an 4:1 \htp:H$^{+}$ ratio. We have been able to duplicate this ratio at higher currents in our own MIST-1 ion source. We expect to keep this ratio and current density, but increase our total power by the use of a new LEBT system, which will be described in the following work.

More information about the MIST-1 ion source can be found in\cite{abs:isodar_cdr1, alonso2021isodar, winklehner:mist1, axani:mist1,winklehner2021high}, and described in full detail in the appendix.

\section{Performance of MIST-1}
The performance of MIST-1 was able to meet the criteria described in section 5.2, with the exception of producing 10 mA of \htp. However, with a larger aperature size and new LEBT design, the current density is sufficient to fufill this requirement. It has confirmed through simulations that increasing the aperature size with the same current density will increase the total current to 10 mA of \htp, while maintaining sufficiently low emittance. More details on the performance of the ion source can be seen in \cite{winklehner2021high} within the appendix. The new LEBT design that will allow the ion source to reach higher currents and match the the RFQ needed for IsoDAR is described in the following section.

\section{Matching Ion source to RFQ}
The LEBT is a series of electrostatic lenses which con-nect the ion source to the RFQ. The electrodes shape and steer the beam into the RFQ to match the desired Twiss parameters at the RFQ entrance. The details of which are located below in this technical document:

\includepdf[pages={-}]{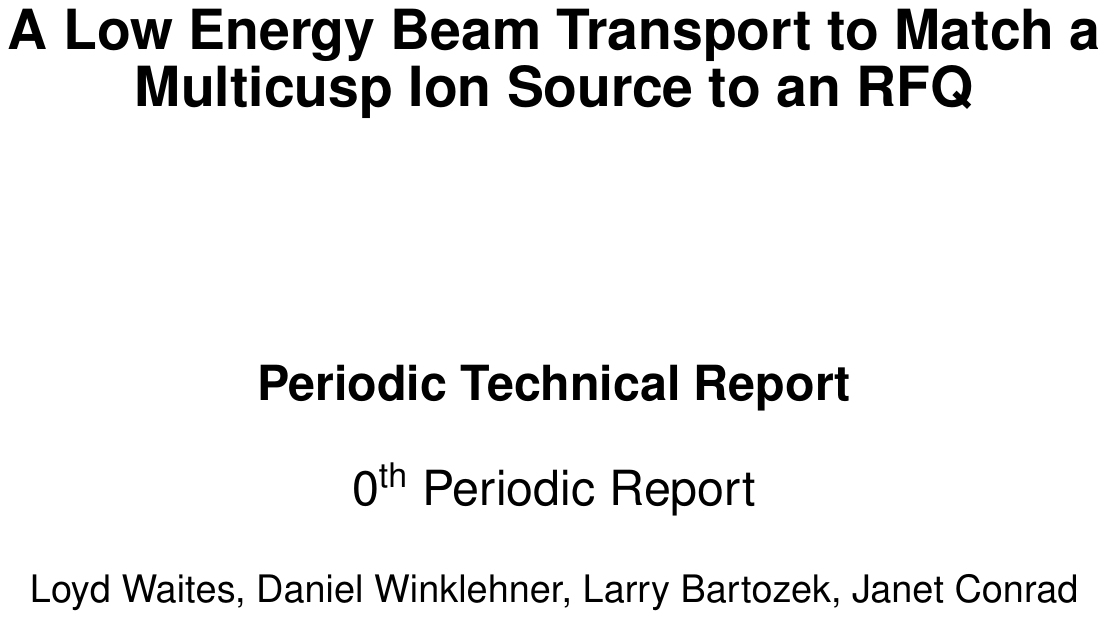}

%% file: chap6.tex
\chapter{RFQ Design}

\section{Physics of an RFQ}
\begin{figure}[t]
   \center{ \includegraphics[width=.7\textwidth]
         {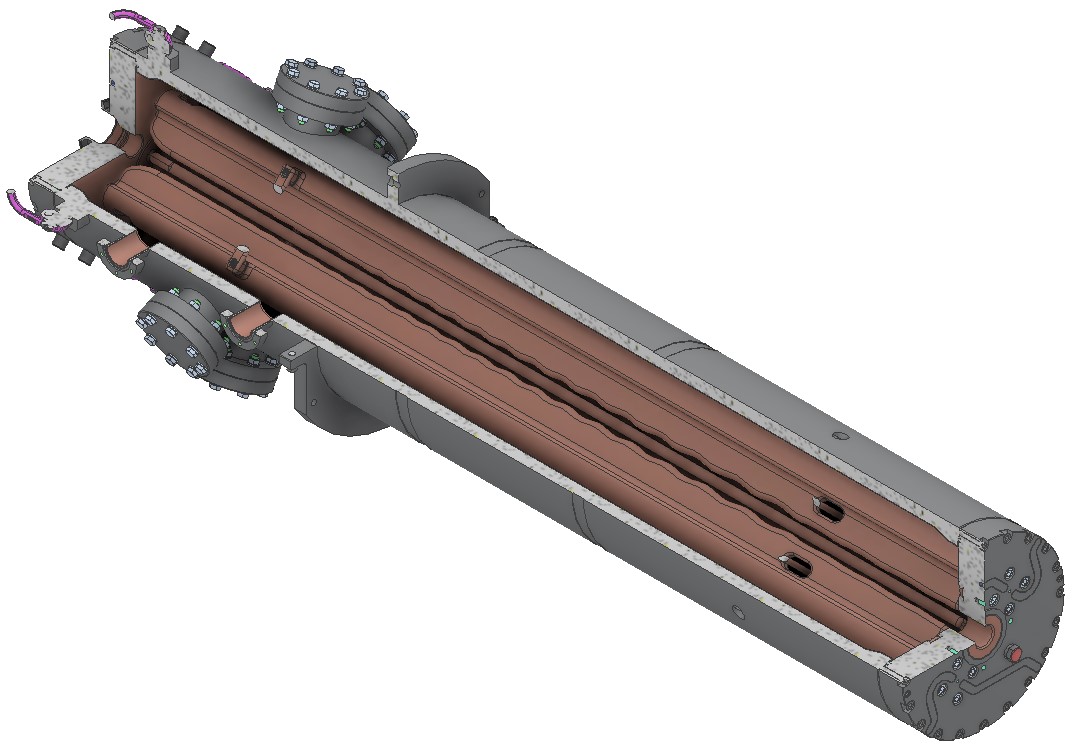}
        \caption{\label{fig:RFQ} 3/4 cut view of the CAD of model for the IsoDAR RFQ, showing the modulated vanes. Beam would be injected in the left side of the figure, and exit through the right. 
         }}

      \end{figure}

Many accelerator systems today use a Radio-Frequency Quadrupole (RFQ) as an early accelerator in the beamline. An RFQ works as an alternating-voltage electric quadrupole that not only accelerates but simultaneously bunches a beam. An RFQ is typically placed following a Low Energy Beam Transport (LEBT) section, which transports the beam from an ion source and matches it to the RFQ.
For high intensity beams at low energy like that which would be used for IsoDAR, space charge is an important factor. Space charge causes a transverse defocusing effect of 1/$\gamma^2$, where $\gamma$ is the Lorentz factor. An RFQ uses its quadrupole fields to focus the beam in the transverse direction while defocusing in the longitudinal direction. This property of RFQs is typically described by the parameter $B_n$ \cite{crandall:rfq}.
While other accelerators use a single cavity and thus accelerate bunches within a specific phase window, RFQs are made up several individual cells that gently shape the beam into individual bunches as it is accelerated. A model of the IsoDAR RFQ can be seen in \figref{fig:RFQ}

The input beam to an RFQ is typically DC. However, the RFQ electrodes have a sinusoidal shape, providing a gentle longitudinal force on the beam in addition to its transverse focusing. This causes the beam to adiabatically bunch. The beam dynamics are thus effected by increasing the space charge forces in the longitudinal direction. The modulation of the electrode varies depending on the longitudinal position of the cell in the RFQ. Later on in the RFQ as the beam gains energy the wavelength of the modulations becomes longer. Each cell has cell length $\ell_c = \beta_c \lambda_{\text{RF}}/2$. This can then be parameterized by each cell $n$ having modulation factor $m_n$. 

The beam is simultaneously bunched and accelerated. The ratio at which this happens is known as the synchronous phase, $\phi_{\text{s},n}$. I.e. if $\phi_{\text{s},n}$, is increased, the beam is accelerated to higher energies. This can be adjusted by the cell lengths. 

We can therefore fully describe the beam dynamics of an RFQ with parameters of Cell Number, $n$; Focusing Strength, $B_n$; Modulation Factor, $m_n$; and Synchronous Phase, $\phi_{\text{s},n}$.

Due to the number of variables being dependent on cell number, this parameter space can be very large. Finding an optimized beam dynamics design can be very time consuming and often takes a large number of design iterations and simulations in various parts of the design space. This is especially true in the case of the IsoDAR, where the design of the RFQ is unconventional due to its requirements of high bunching and low acceleration. This means mostly unknown design spaces must be explored.

\section{RFQ Design using Standard tools}

\begin{figure}[tb]
        \center{\includegraphics[width=.6\textwidth]
         {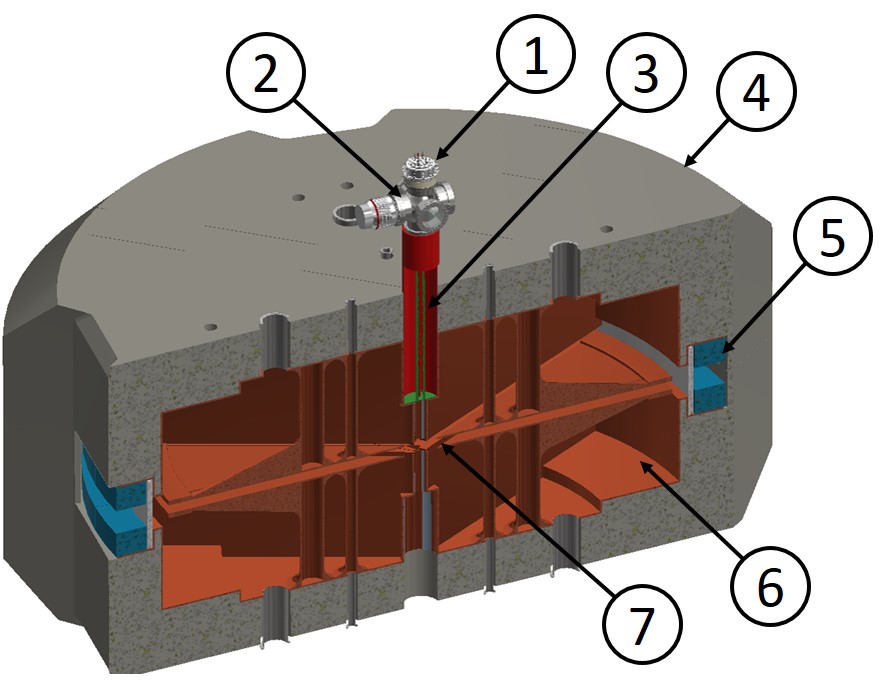}
         }
        \caption{\label{fig:accelerator cartoon} {Conceptual layout of the compact cyclotron driver for high intensity
                 60 MeV/amu \htp beams.
                 1. Ion source,
                 2. diagnostic and pumping box,
                 3. Radio Frequency Quadrupole (RFQ) injector,
                 4. Compact cyclotron yoke,
                 5. Compact cyclotron coils,
                 6. Accelerating RF electrodes (``Dees''),
                 7. Central region.}  \cite{winklehner:nima} }
                 
 \end{figure}

RFQs can be used at low energies as a pre-accelerator before being injected into a larger accelerator. Due to the IsoDAR experiment taking place underground, compactness of the cyclotron system is of utmost importance. To achieve this goal, an RFQ takes the place of a longer LEBT injection system. The RFQ must also axially inject the beam to prevent beam spread the RFQ must be placed inside the cyclotron. This acts to further compact the system, as well as greatly increase end to end transmission. A CAD rendering of the RFQ injected IsoDAR cyclotron can be seen in Figure \ref{fig:accelerator cartoon}.

The RFQ design parameters are often determined by the cyclotron geometry and properties. For example, length and diameter of the RFQ are determined by the size of the iron yoke in the cyclotron's central region. The RFQ was designed to be split coaxial due to its small diameter and low frequency requirement. The output energy of the RFQ (70 KeV) is based on the maximum allowable energy that can be transported by the spiral inflector. The input energy (15 KeV) is set to minimize the length of the RFQ while maintaining its frequency.

\begin{figure}[t]
        \center{\includegraphics[width=.7\textwidth]
         {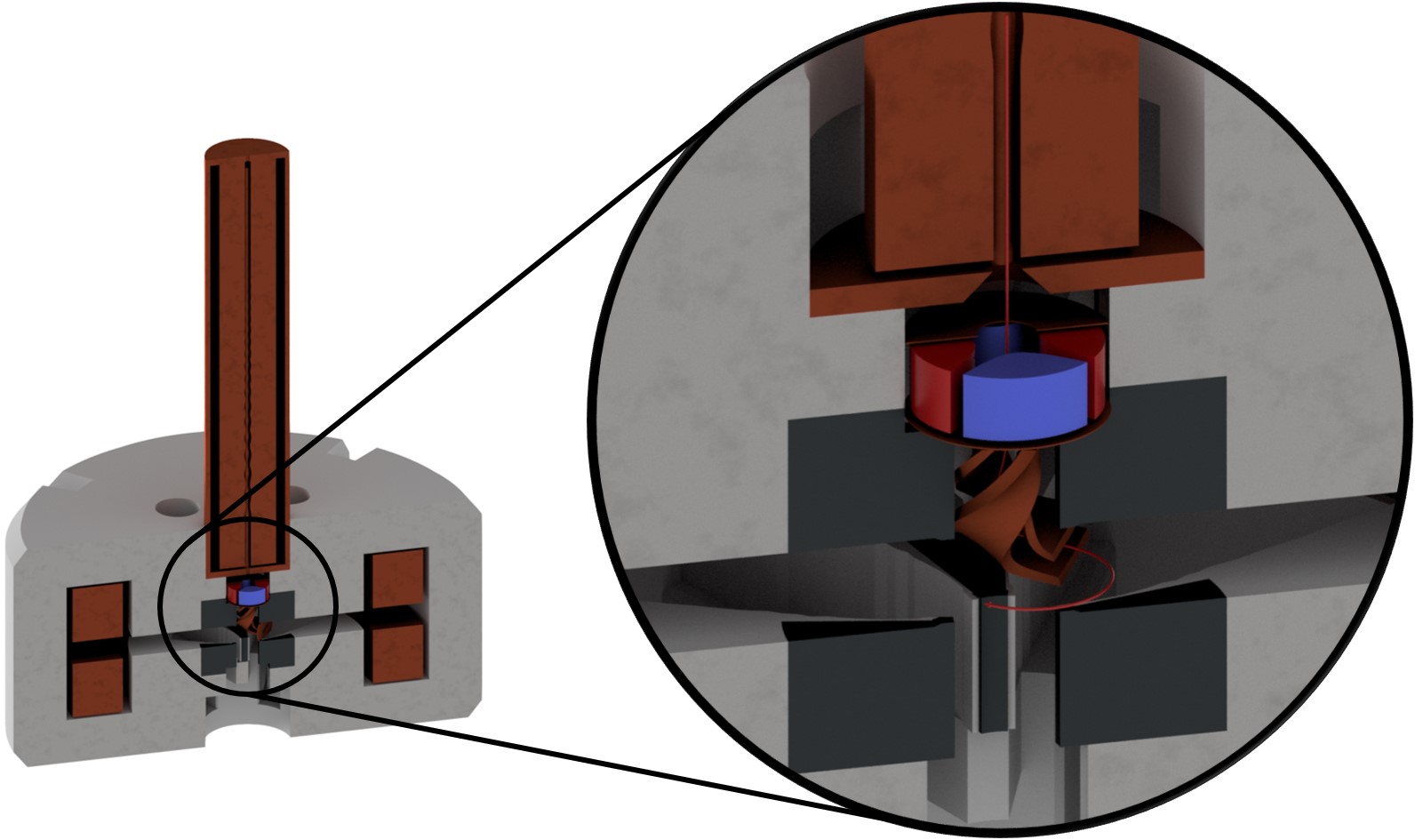}
         }
        \caption{\label{fig:central_regionl} Cross sectional view of the cyclotron central region with RFQ, rebuncher cell, and spiral inflector.
\cite{winklehner:nima}
 }
        
      \end{figure}

The RFQ injects a bunched beam which is in sync with the phase of the cyclotron frequency of 32.8 MHz. The beam is injected from the RFQ into the cyclotron axially; however, this is perpendicular to the accelerator plane in the cyclotron. After leaving the RFQ the beam is injected into a rebunching cell, reducing the energy spread. The highly bunched beam is then fed into a spiral inflector, which changes the direction of the beam. A CAD model of the IsoDAR spiral inflector can be seen in Figure \ref{fig:central_regionl}

The IsoDAR RFQ is used primarily as a pre-buncher for the 5~mA \htp beam which will be axially injected into the IsoDAR cyclotron \cite{winklehner:rfq1}. This design has been investigated and optimized using traditional simulation techniques, as well as surrogate models from machine learning. 

Time consuming simulations are mitigated by using fast RFQ dynamics codes such as RFQGen and PARMTEQM, however these codes are not always accurate, and still time consuming when large amounts of simulations are required for a full optimization. More accurate particle-in-cell codes are extremely computationally expensive and are not ideal for use of design optimization

\section{Machine Learning Models for RFQ Optimization}
Surrogate models are used in place of more complex simulations in order to reduce computation time. This has been shown to be a useful tool in for particle accelerators \cite{PhysRevAccelBeams.23.044601}. 

The data generated to use surrogate models and machine learning is of course different depending on its intended use. For our studies, we investigated two cases. Creating a real-time tuning tool for RFQ beam dynamics, and creating a tool to confine the design parameter space of an RFQ.

The first case is based on using a fixed  RFQ design. This surrogate model takes a set of values describing the beam dynamics at the entrance to the RFQ, then simulates the final output properties of the beam at the end of the RFQ. This can thus be used for tuning the input parameters to the RFQ from the LEBT in real time, and help us match our simulated and experimental data.

The second case uses the full description of an RFQ as well as the input beam parameters to restrict the deign space.

\paragraph{ Neural Networks}
Neural Networks are a biologically-inspired programming structure that transform data over a series of interconnected nodes. Each node is in a layer in which the nodes from the previous layer may be connected to it. In a fully connected neural network, as I will be describing for this work, each node in each layer is connected to each node in the previous layer. Each node uses a linear transformation, or 

\begin{equation} \label{linear_fit}
T(x):=Wx+b
\end{equation} 

 where $W=(a_{ij})\in \mathbb{R}^{m\times n}$, $x\in \mathbb{R}^{n}$, $b\in \mathbb{R}^m$, and $n,m \in \mathbb{N}$. $W$ and $b$ known as the weights and biases of the neural network. The node also uses an activation function, which defines the output of the node based on its input. This determines the information passed to the next layer.
 
 In the most simple terms, the activation function could be binary. This often describes a classifier, ie. the subject of an image either is or is not a cat. The three most common activation functions that are used are 
\begin{itemize}
    \item Sigmoid ($\frac{1}{e^{-x}}$)
    \item Hyperbolic Tangent
    \item Rectified Linear Unit (ReLU), which returns a value of zero or a value on a linear scale.
\end{itemize}

\begin{figure}[t]
        \center{\includegraphics[width=.7\textwidth]
         {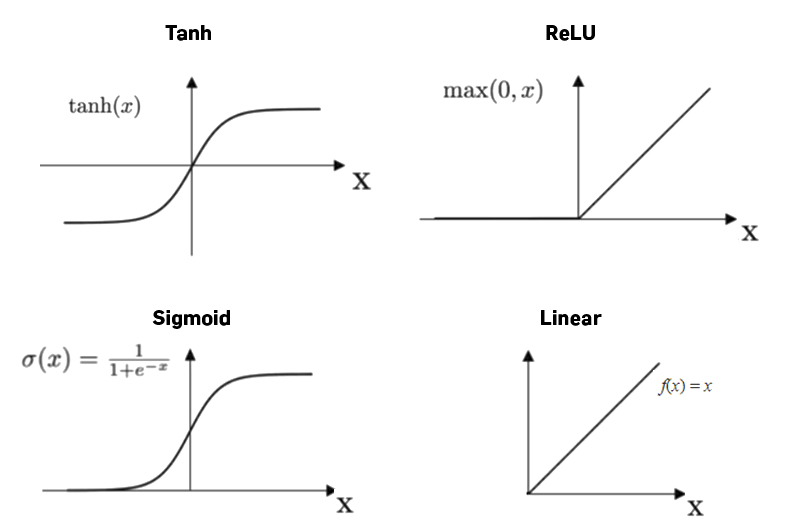}
         }
        \caption{\label{fig:acitivation_functions} Graphs showing different activation functions of tanh, signmoid, ReLU and linear. From
\cite{aiwiki}
 }
        
      \end{figure}

Each of these activation functions have a distinct purpose. Sigmoids have the best sensitivity around their center, but also have an issue with saturation. Very large or very small values converge slowly because the gradient in the sigmoid function's gradient decreases at the extremes. A hyperbolic tangent function has a stronger gradient than a sigmoid, however it still has issues with a vanishing gradient. A ReLu function does not have a vanishing gradient due to its linear nature, and is typically the  most common, but are not differentiable at zero and so are not appropriate for use in recurrent neural nets. Linear activation functions are also occasionally used, however they do are not able to model non-linear functions, and so are often replaced by ReLu. For our case, we have used ReLu functions. These activation functions can be seen in Figure \ref{fig:acitivation_functions}.

Combing the weights, biases, and activation function provides a continuous function which can be used to transform data. When a neural network is "trained" it is given a series of input data which is matched to a set of output data. The dimension of these data sets must match the first and last layer of the neural network, respectively. The neural net then modifies the its weights and biases (which are each a matrix of values with the shape of the neural network)  from equation \ref{linear_fit} in order to transform the input data to produce the output data. 

The weights and biases are randomly initialized and then tuned using an optimizer, often some form of gradient descent. More recently, the Adam optimizer has been used rather than classic stochastic gradient descent. Adam uses an Adaptive Gradient Algorithm  as well as Root Mean Square propagation in order to make the optimization, and thus learning time, more efficient. Due to this, we have chosen to use the Adam optimization the training of our neural network.

\begin{figure}[t]
        \center{\includegraphics[width=.7\textwidth]
         {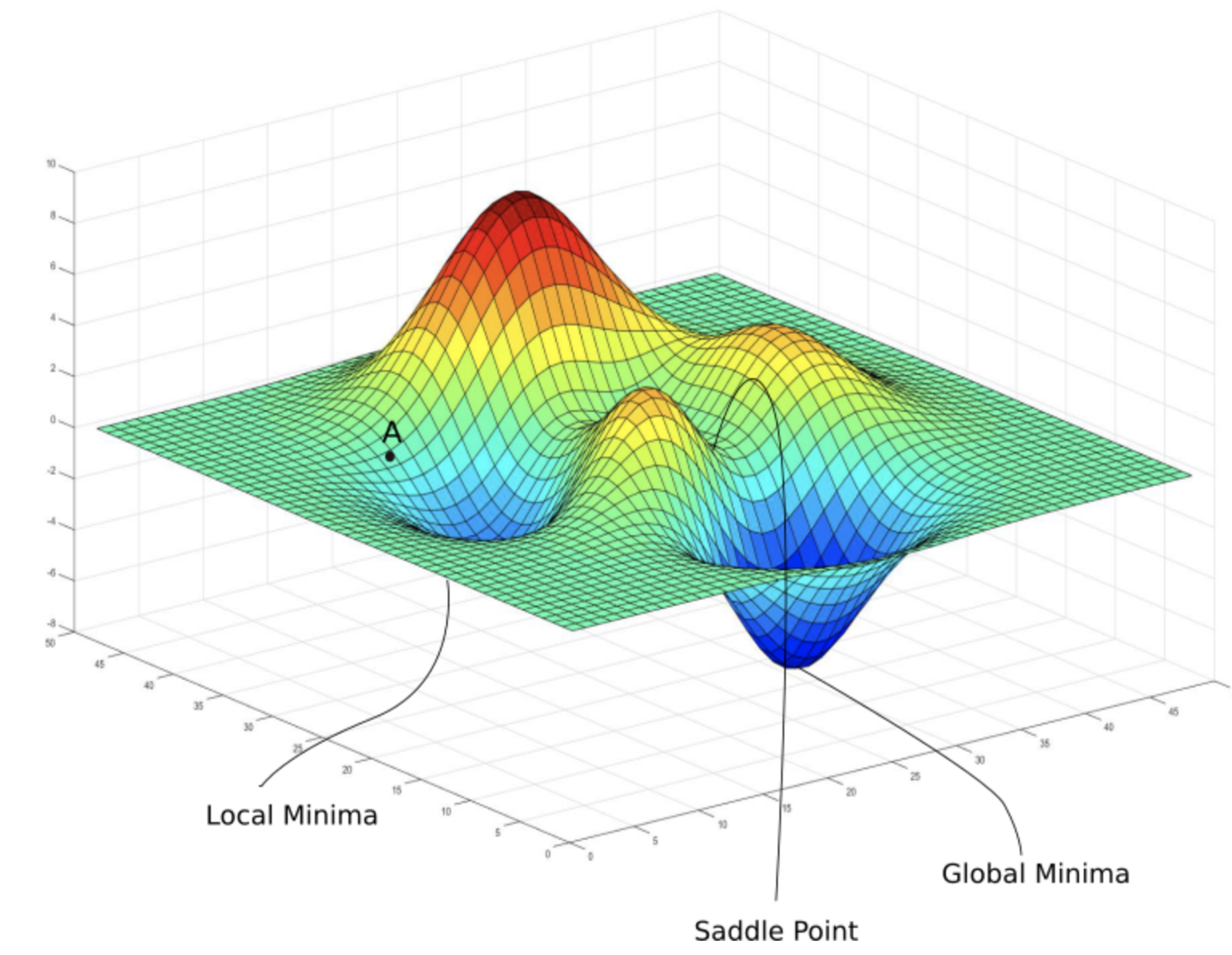}
         }
        \caption{\label{fig:manifold} a 3D manifold showing local and global extrema. From
\cite{manifold}
 }
 \end{figure}

While there is some concern about the optimizer finding a local extrema rather than a global extrema, the often very high dimensionality of problems typically used for machine learning prevents this. The higher the dimensionality, the more complex the parameter manifold is, leading to it being less likely that the optimizer will be caught in a local extrema. This is visualized in Figure \ref{fig:manifold}.

Neural networks can be evaluated based on the error in which they reproduce output data through the transformation described above. Once data is generated, the data is randomly separated into a training and test set. The data can first be compared to the data it was trained on to investigate a reasonable reproduction of the training data set. However, the trained neural net also must be tested on data of the same structure that it was not trained on. If the error of test data set is significantly larger than that of the the train data set, this is indicative of over fitting the training data. To prevent this, larger data sets or tools such as dropout, when individual nodes are deactivated during training to prevent dependence, may be used. 

An untrained neural network can be described by its hyperparameters, such as the number of layers or "depth," number of nodes in each layer, activation functions of each node, number of epochs or passes, and connectedness. To ensure the best neural network shape, a second degree of optimization can be used on entire neural nets. This is called hyperparameter optimization. Neural networks are trained on a given data set and then compared to some metric (in our case, the error between the original and reproduced output data set). The hyperparameters are then altered, creating a new shape for the neural network, and it is trained again. The hyperparameters are changed until either there have been sufficient iterations, or the error has been deemed sufficiently low by the optimizer. 

Many of these tools have been made accessible by programs like tensorflow \cite{tensorflow2015-whitepaper} and keras\cite{chollet2015keras}. These packages provide a simple interface in python which allows for determining the shape of a neural network, its hyperparameters, and provides tools for hyperparameter optimization. These were used in our work to find a neural network to generate a surrogate model of an RFQ.

Neural networks have been used for approximations and general surrogate models in past works \cite{Cybenko1989,HornikSW89}. Authors were able to examine the similarity of surrogate models to real data in a single hidden layer neural network. This was later expanded to error estimates based on the full set of hyperparameters. This is reviewed in ~\cite{ellacott_1994} and~\cite{pinkus_1999}. How these have been used for accelerators has been described in detail in the attached work.

\paragraph{Polynomial Chaos Expansion}
Uncertainty quantification has also been used in creating surrogate models for accelerators \cite{adelmann-2019-1}. 

A polynomial chaos expansion (PCE) is another form of machine learning that uses superpositions of polynomials in order to estimate a complex function. With some input data x, a model can be constructed using orthogonal functions:

\begin{equation}
    m({x}) \approx \hat{m}(x) = \sum_{i=1}^{P}c_{i}\Psi_{i}({\xi})
                                      = \sum_{i=1}^{P}c_{i}\prod_{j=1}^{d}\psi_{j}(\xi_j)
\end{equation}

where

\begin{equation}
    P = \frac{(p+d)!}{p!d!}
\end{equation}

 $d$ is the dimensionality, $p$ is the order of the polynomials, $c_{i}$ are the expansion coefficients, $\psi_{i}$ are the orthogonal multivariate polynomials and $\psi_{j}$ are the univariate polynomials, where j is the dimension that corresponds to the dimension of the input. I.e. Hermite polynomials correspond to normally distributed dimensions, where uniformly distributed dimensions correspond to Legendre polynomials. ${\xi}$ describes the input data vector. 

Similar to the weights in a neural network, PCEs work by altering the expansion coefficients, $c_{i}$,  to provide the best possible mapping from import to output vectors. The three most common ways to do this is by regression or Bayesian methods.  each of which are flexible, however should have a sufficient number of samples to be accurate \cite{SUDRET2008964}.
    
    \begin{equation*}
    N = (d - 1) P.
    \end{equation*}
    
Alternatively, orthogonal projection can be used. in which the training data has an exponential growth with dimension. I.e.                 \begin{equation*}
    N = (p+1)^{d}.
    \end{equation*}

A large advantage of using a PCE as a technique in ML is the use of Sobol indices \cite{Sobol01}. Sobol indices are able to describe the change in global output based on a change in each individual input variable. This is a useful tool to constrain variables who's effect on the final output variables is minimal, while focusing on variables who's effects are more significant. The total sensitivity describes the effect of an input variable on the set of output variables, including those which are correlated with other dimensions.

For more information on PCE a useful resource is \cite{adelmann-2019-1}. 

Our full description of how we used machine learning to design and optimize our RFQ can be seen here:
\includepdf[pages={-}]{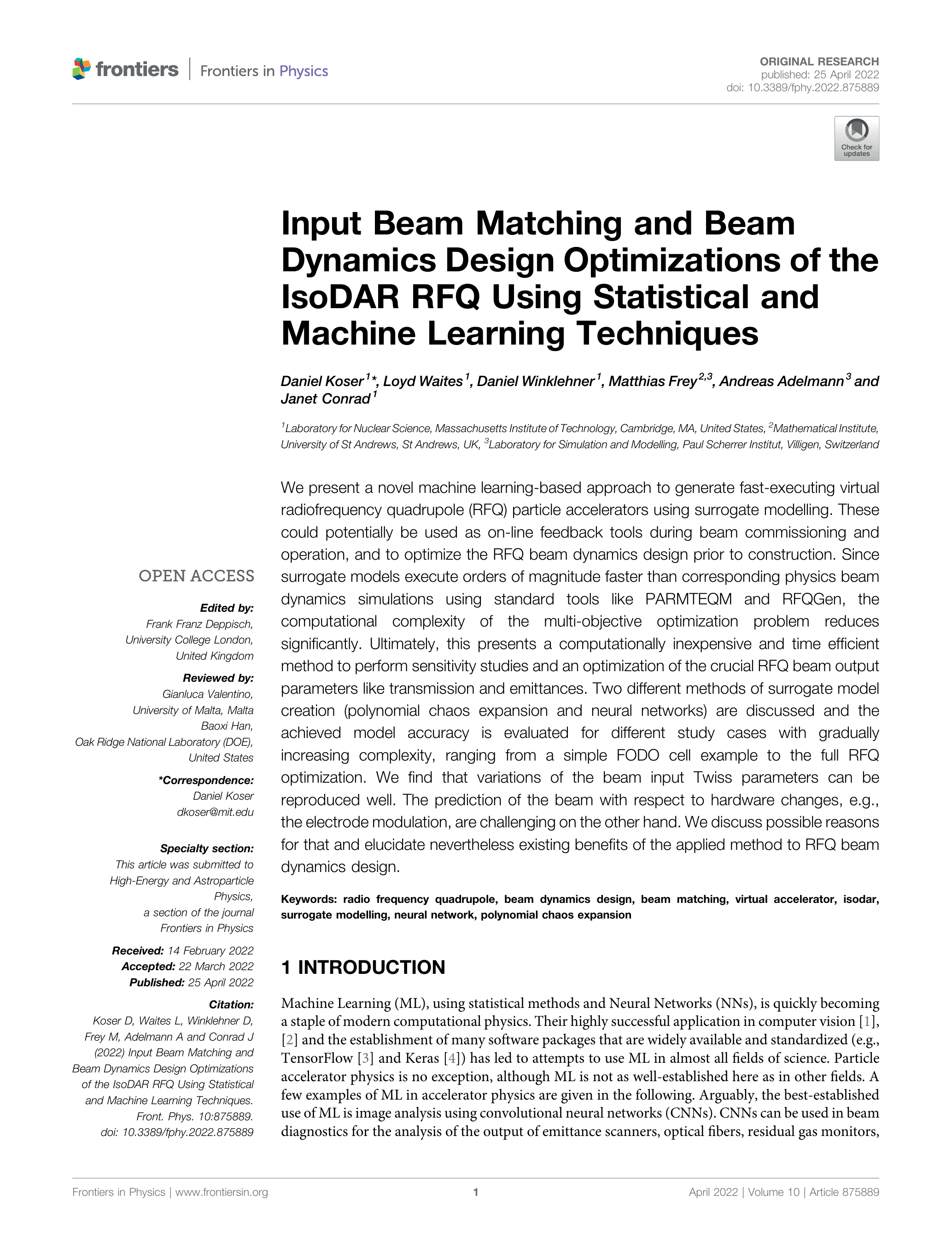}

%% file: chap7.tex
\chapter{Medical Isotope Production Using the IsoDAR cyclotron}
The IsoDAR experiment was designed for a neutrino experiment. However, the requirements of this ambitious neutrino experiment have pushed the limits of cyclotron technology. The most common use for cyclotrons is the production of radiopharmaceutical isotopes. The high currents provided by the IsoDAR cyclotron are well suited to create high yields of isotopes with relatively small cross sections or long half lives, as the high power provides more protons on target to produce these rare isotopes. Due to its high power it is also able to produce these high yields at a rapid rate.

The prohibitive cost of production of these rare isotopes is a bottleneck that prevents revolutionary treatment from being used in clinic. While the DOE and some commercial groups have been increasing their production of rare isotopes, there is a growing unmet demand.

The IsoDAR cyclotron can also be easily reconfigured to accelerate other ions with a charge to mass ratio of 2. For example, the magnetic fields may be tuned to accelerate $\sim$2 mA of alpha particles up to 240 MeV, with no physical changes to the cyclotron. This beam species variability opens up the potential for the IsoDAR beam to multiple new nuclear reactions of interest. 

\section{State of Medical Isotope Production}
Radiopharmaceuticals are being increasingly used in medicine. They have been used for diagnostic practices such as imaging \cite{townsend2004physical,rahmim2013resolution,galldiks2019current,de1998radioisotopes,schmor2011review}, and therapeutic applications such as cancer treatment, and even treatment of HIV \cite{schmor2011review,milenic2004targeting,rosenshein1983radioisotopes,finlay2005radioisotopes,garg2020comparison}.

\begin{figure}[!htb]
    \center{\includegraphics[width=.5\textwidth]
    {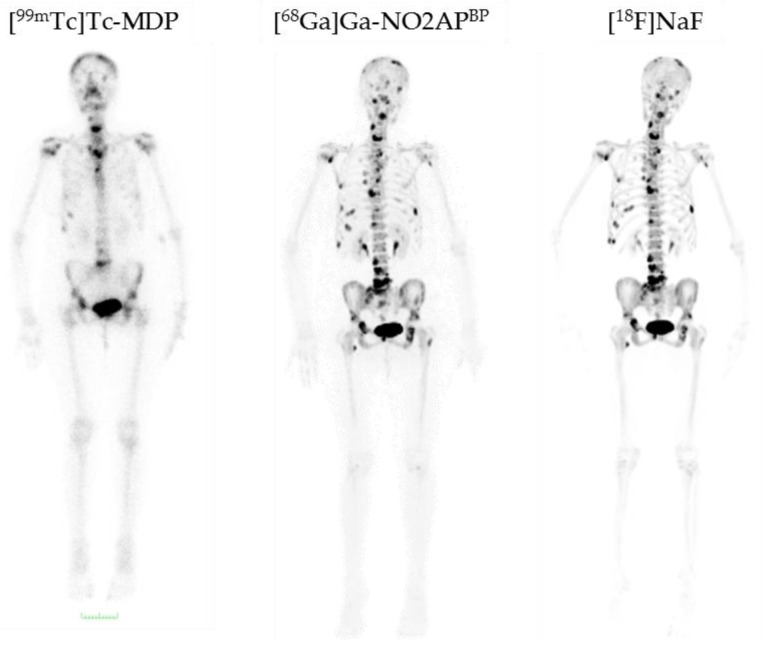}}
    \caption{\label{fig:pet} Imaging comparison of three different medical isotopes. The left most is a SPECT scan using $^{99m}$Tc, the daughter of $^{99}$Mo. The next use $^{68}$Ga, and $^{18}$F, both of which are PET isotopes. The PET isotopes show much higher resolution, and more tumors are visible. Taken from  \cite{pfannkuchen2017novel}.}
\end{figure}

In the case of PET imaging, a radioisotope is attached to a tracer which then seeks out a binding site, such as a tumor or neurotransmitter binding site, and then radioactively decays releasing a positron. The positron then annihilates with the electrons in the immediate region producing two 511 KeV gamma rays. These gamma rays are emitted in opposite directions, creating a detectable signal from a PET scanner. This provides precise position resolution where the isotope decayed, and thus what tissue the tracer had bound to. SPECT scanners, known for using Mo$^{99},$ are able to achieve similar results by measuring the gamma ray from a radioisotope rather than the annihilation signal from positron emission. While these processes are cheaper, they are lower resolution than PET scans, and can even sometimes miss signal entirely, as seen in \figref{fig:pet}. As PET scanners decrease in cost, the demand for higher resolution PET imaging, and thus the isotopes it requires, is expected to grow.

\begin{figure}[!htb]
    \center{\includegraphics[width=.8\textwidth]
    {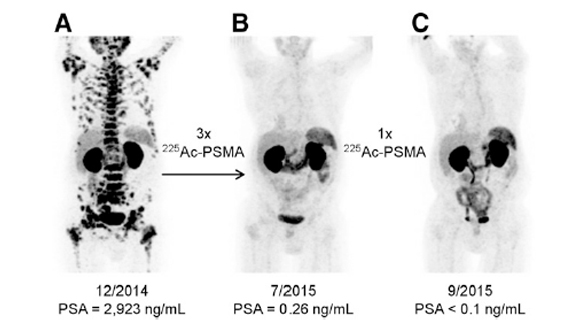}}
    \caption{\label{fig:ac225} PET images over time of a subject with malignant cancer. After two treatments with $^{225}$Ac the patient has become essentially cancer-free. Taken from  \cite{kratochwil2016225ac}.}
\end{figure}

Using radioisotopes for treatment is a broad space. The concept is similar to a PET isotope- attach a radioisotope to a tracer which binds to a targeted region. In this case, the goal of the radioisotope is not to provide an image, but instead cause localized cell death. Any form in which radiation from a targeted radioisotope can cause cell death can be used as a form of treatment to kill of unwanted cells (such as tumors). This can be done in many forms, however for the sake of this thesis, I will be focusing on alpha emitters. The advantage of alpha emitters are their highly localized area of effect that result from the rapid loss of energy of an alpha particle in matter. This is the ideal property of a radioisotope treatment- first, high amounts of energy ensure a double stranded break of the cell, causing death rather than mutation. Second, the energy is deposited over a small area, preventing damage to any neighboring cells. This has lead to the exciting results of alpha emitters in recent years that can successfully treat even malignant tumors, such as those seen in \figref{fig:ac225}

There is a delicate balancing act when deciding on an isotope for treatment or therapy. On one hand the half life cannot be so short that the isotope decays before reaching its target. On the other, the isotope half life can not be so long as to over expose the patient after the therapy is completed, or not provide a sufficiently strong signal. The chemical nature of the radiotracer pair must also be made to ensure that the substance and its products is not toxic. 

One way some of these challenges have been mitigated is by the use of "mother" and "daughter" isotopes. In some cases, such as with $^{99}$Mo, a relatively long lived  isotope is produced at an accelerator facility, then shipped to a hospital. The long lived isotope is then "milked" to its daughter isotope, or $^{99m}$Tc, which is short lived, and is responsible for imaging. This allows isotopes to be transported long distances, while simultaneously allowing a strong signal with a different isotope. In this regime, it is preferred to have a long lived mother isotope to eliminate a rapid supply chain, and a short lived daughter isotope to provide strong signal. This is another advantage of several PET isotopes over SPECT, as many PET isotopes  follow this model.  

A new growing field of interest is using "theranostics," or a combination of radioisotopes to simultaneously and in the same location both image and cause treatment. In some cases this is done with a single isotope \cite{kelkar2011theranostics,svenson2013theranostics,yordanova2017theranostics,langbein2019future,del2007nuclear, weber2020future}.  This would be a case of an isotope that in its decay chain emits both a detectable decay element (photons in the case of SPECT, or positrons in the case of PET). While also having therapeutic ionizing radiation that can kill malignant cells or biological elements in the area.  In other cases it is done with two separate isotopes with similar chemical properties which allow them to bind to the same ligand or protein \cite{bailey2021developing}. One way to do this is to produce two radioisotopes with similar chemical properties, ensuring a general case that they would be able to bind to the same tracer, regardless of application. Because the diagnostic and therapeutic are bound to the same carrier, the tracer will be bound in the same location, therefore tracking the therapeutic and ensuring healthy tissue is not damaged. 

These medical tools are based in the production of unstable isotopes. The primary ways this can be done is in a nuclear reactor or by using an accelerator. Accelerators have become more common as they are a more controlled, reliable, and cleaner way to produce mass amounts of desirable  isotopes. While linacs are used by some groups, cyclotrons are much more common due to their lower cost, higher compactness, and capability for redundancy.

Most commercial cyclotrons accelerate $\sim$1 mA of H$^-$ ions to between 30 and 70 MeV \cite{schmor2011review}. This energy range is useful for producing a wide range of isotopes, including $^{99}$Mo at lower energies, and $^{134}$Ge,  and $^{225}$Ac  at higher energies \cite{weidner2012proton}.

The limitations on these isotopes are their low production rates. This is due to bounds in beam current and thermal constraints on production targets. IsoDAR is capable of mitigating both of these bottlenecks. The high current of IsoDAR provides more protons on target allowing more isotopes to be produced. IsoDAR is also different than other cyclotrons in its extraction of \htp. \htp has the advantage of easily being broken off, providing for development of higher power targets, as will be described later in this chapter.  

\section{ Introduction to the use of IsoDAR to Address Medical Isotope Demand}
\includepdf[pages={-}]{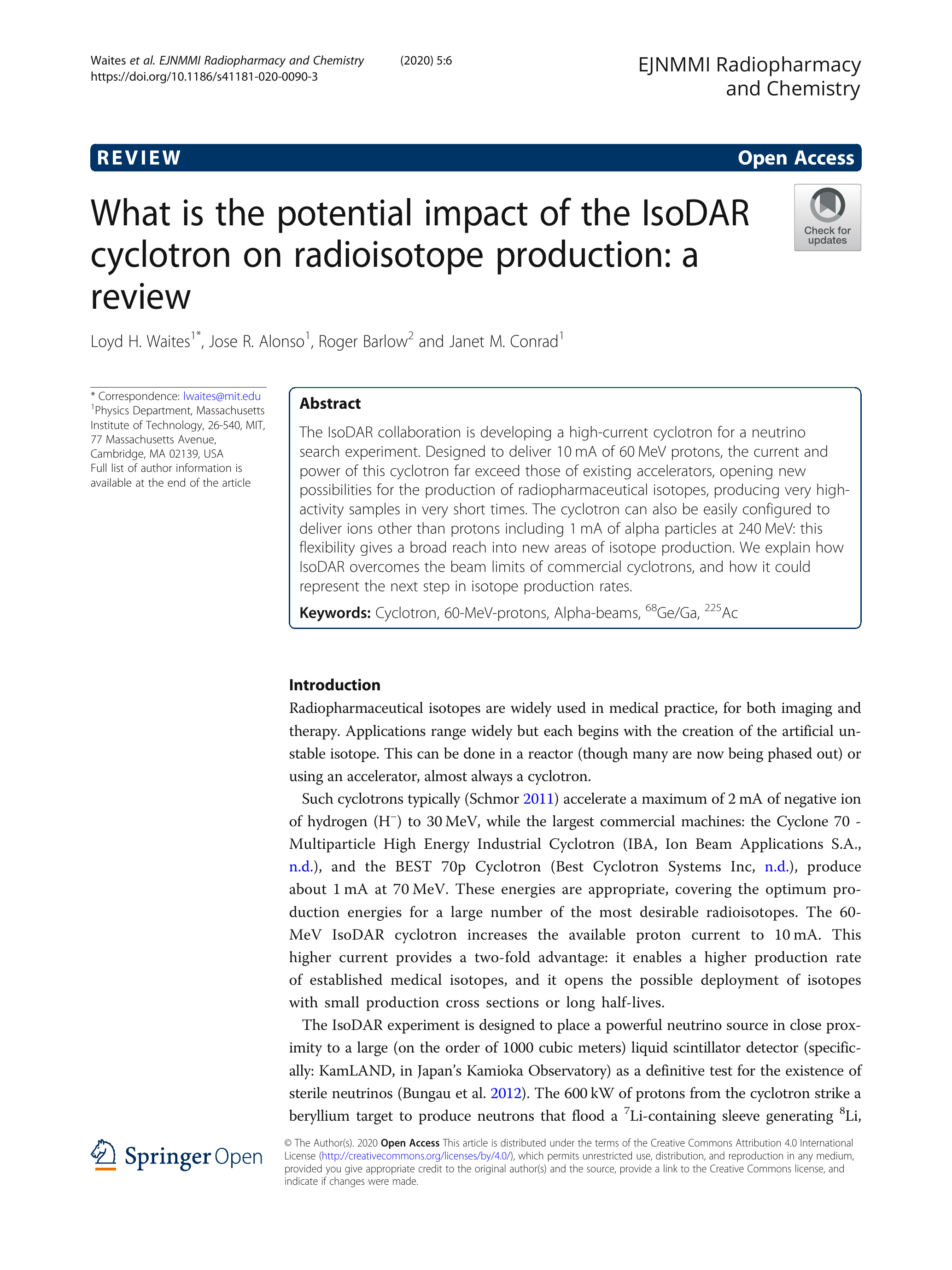}

\section{Further Information at applying IsoDAR to Isotope production}
 Research and development on improving high power targets has been an ongoing effort at labs sponsored by DOE, however progress is incremental. To increase progress in this regard, we propose using the IsoDAR cyclotron as a testbed for target development. While the total power of the IsoDAR cyclotron overwhelms the power capabilities of modern production targets, the \htp beam extracted from the IsoDAR cyclotron can be split into multiple lower power beams, each with individually variable power. The beam can be distributed among several stations, each with different requirements for isotope production or nuclear experimentation. This use of stations is expanded on in the work below.

\includepdf[pages={-}]{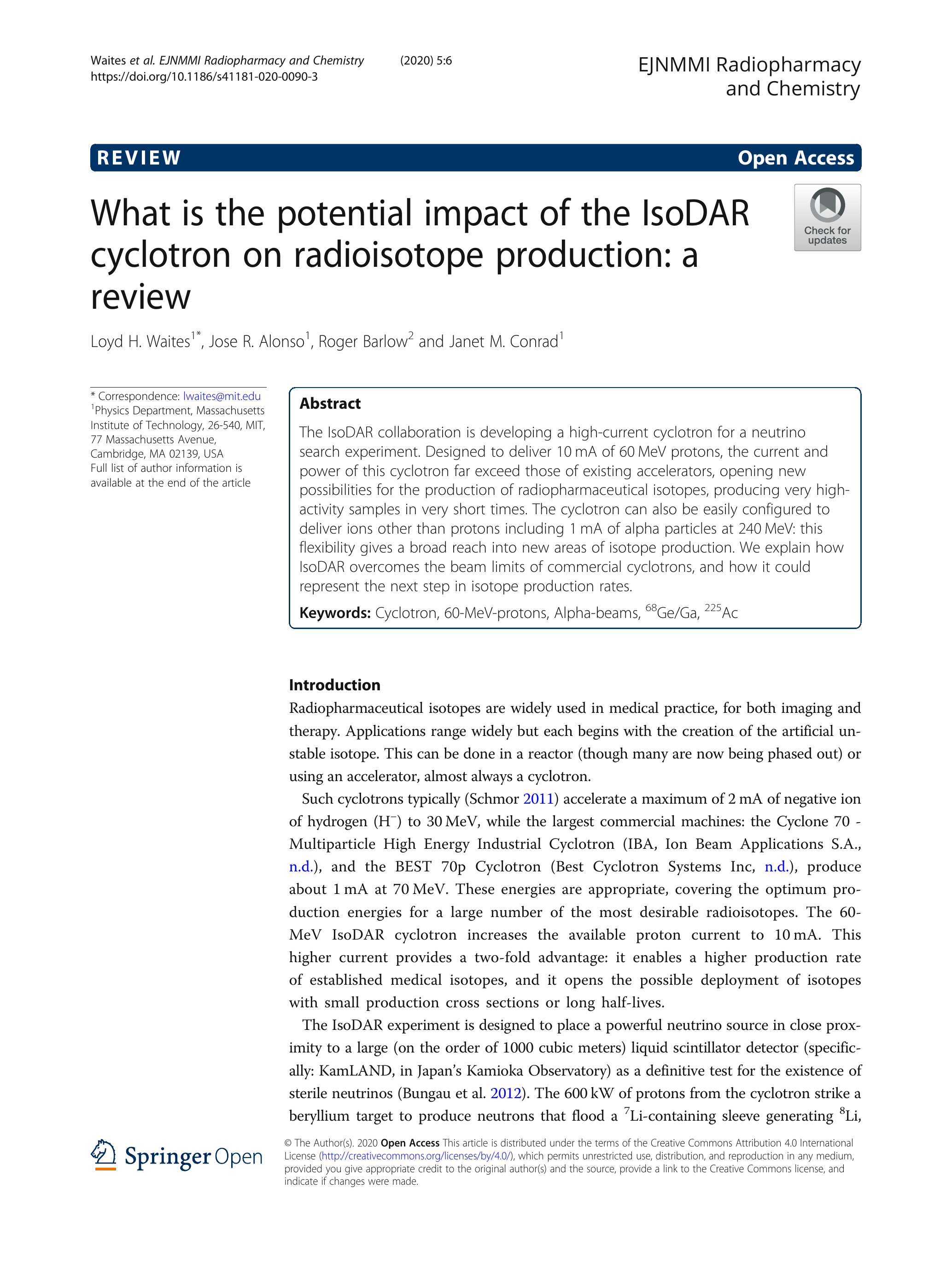}

\section{Potential for Expanding The Design}

As mentioned previously, it is possible to use the same IsoDAR cyclotron on other ions with a mass/charge ratio of 2. We have investigated this process with alpha particles, in which we have found that by tuning the magnetic field to compensate for the slight change in mass, with no changes to the magnet shape, it is possible to accelerate alpha particles along the same path as \htp ions. Because alpha ion sources do not supply as high current as our \htp ion source, issues from space charge are further mitigated. One additional feature of using alpha particles over \htp is that the acceleration of the cyclotron is 60 MeV per nucleon, meaning that at extraction the total energy of the alpha beam would be 240 MeV. This unlocks new reactions at these high energies. 

Using GEANT4 we executed a preliminary study of isotope production with these 240 MeV alphas. We found two cases of particular interest.

We investigated the production of $^{225}$Ac using these higher energy alphas compared for our previous estimates of 60 MeV protons. While we found that the yields per unit current using 240 MeV alpha particles on a radium target were lower than the yield of 60 MeV protons (and lower per unit current than the 100 MeV protons used by DOE) on a thorium target, we found that there was a higher ratio of $^{225}$Ac as compared to $^{227}$Ac when produced with 240 MeV alphas. $^{227}$Ac is a common contaminant that is often found in conjunction with $^{225}$Ac. This is concerning due to its long half life, and similar chemistry, leading to radioactive material existing in the patient even long after treatment, potentially causing damage. The effects of the contaminant are still under investigation, however producing higher purity $^{225}$Ac is certainly a favorable outcome \cite{campbell1956metabolism,baimukhanova2022alternative, abergel2019actinium, engle2018production, miederer2008realizing}. Based on our GEANT4 simulations 100 $\mu$A of 100 MeV protons on a thorium target for 1 hour as done by DOE would produce $\sim$230 mCi of $^{225}$Ac and 0.51 mCi of $^{227}$Ac, or a ratio of 451:1. Alternatively, using 100 $\mu$A of 240 MeV alpha particles on a radium target for 1 hour would produce $\sim$73 mCi of $^{225}$Ac and .047 mCi of $^{227}$Ac giving a ratio of 1550:1. While the yield of $^{225}$Ac is lower per unit current in this case, as described previously in this section this can be overcome by the higher total current capability of the IsoDAR cyclotron. Therefore, using this reaction will allow for high production of higher purity $^{225}$Ac if the power of the IsoDAR cyclotron is properly managed.

The second case was the production of $^{134}$Ce. $^{134}$Ce is a PET isotope that has similar chemical properties to $^{225}$Ac, allowing them to bind to the same tracer and act as a theranostic pair \cite{bailey2021developing}. Being able to simultaneously produce $^{225}$Ac and $^{134}$Ce with the same machine would allow for a "one-stop-shop" radioisotope cancer treatment, in which the $^{225}$Ac would be able to eliminate the tumor with alpha therapy, and the $^{134}$Ce would be able to track the progression of the treatment and the location of any $^{225}$Ac being used in the therapy.

Based on similar GEANT4 simulations, 100$\mu$A of 240 MeV alphas on a $^{140}$Ce target for one hour would produce $\sim$117 mCi of $^{134}$Ce, while 100$\mu$A of 100 MeV protons on a natural Lanthanum target as practiced by DOE produces $\sim$100 mCi \cite{bailey2021developing}. This would indicate that the production of $^{134}$Ce could be made more efficient by the use of high energy alpha particles, which could then be compounded on by the higher current of the IsoDAR cyclotron.

It should be noted that these studies are preliminary. While comparison of the GEANT4 estimates to experimental proton reactions are readily available for benchmarking, the libraries used for the alpha reactions do not have a readily available experiment to directly compare to. Further investigations of the libraries used in GEANT4 should be done before making any final conclusions. 

While the IsoDAR cyclotron is nominally designed to be used at 60 MeV, the design can be modified to create a full family of cyclotrons that exist at various energies. This allows the advantages of higher currents and an \htp at any beam energy that may be useful for an experiment or isotope production facility. For example, an IsoDAR 30 MeV machine could compete with and IBA C-30, acting as a workhorse of $^{99}$Mo production with its 10X higher current. Or the cyclotron could be made larger, increasing the energy to 70 MeV, or perhaps even 100 MeV, thus out compete the IBA C-70 or even the DOE linear accelerators. The details of the design of this "family of cyclotrons" is forthcoming, but we are optimistic of their capabilities.

%% file: Conclusions.tex
\chapter{Conclusions}

The IsoDAR cyclotron is a new tool for understanding physics and engineering applications. In this thesis, I described IsoDAR's potential impact on fundamental physics and its design.

The IsoDAR experiment was originally motivated as a sterile neutrino experiment. I have discussed why sterile neutrinos are a strongly motivated particle in which to search for. IsoDAR would set high confidence, world leading limits on sterile neutrinos which definitively cover preferred regions found in global fits. IsoDAR is also one of the few experiments capable of evaluating models with several sterile neutrinos, evaluating not only 3+1, but also 3+2 and 3+1+decay \cite{alonso2022neutrino}. 

In addition, the experimental setup of IsoDAR leads to the production of a large number of photons as a  convenient byproduct. These photons are primarily produced via nuclear de-exitations, and can be used in conjunction with the Yemilab detector to execute an axion search. This would also set world-leading limits in several axion models \cite{waites2022axion}. 

I outlined the physics of cyclotrons, and the design of the IsoDAR cyclotron. IsoDAR using three major breakthroughs to achieve its higher currents- the use of \htp, RFQ direct injection, and the utilization of vortex motion. Our simulations show that our cyclotron is capable of accelerating and extracting 5 mA of \htp with acceptable loses on the extraction septum \cite{alonso:isodar_cdr2}. 

I described the design and commissioning of our \htp ion source and LEBT electrode system. Our ion source creates a high current, low emittance beam with 80\% \htp  \cite{winklehner2021high}. Our LEBT is designed to properly match the ion source into our RFQ, allowing for >90\% transmission, while providing beam diagnostics and machine protection.

I also described our RFQ-DIP system, its importance to the cyclotron community, and how we used machine learning to optimize its design. The RFQ bunches the beam so that the frequency of the bunches match the frequency of the cyclotron, greatly increasing the end-to-end transmission \cite{winklehner:nima}. We also optimized this design using machine learning, and produced useful real time tools for RFQ commissioning \cite{koser2022input}. 

I closed with the most direct application of the IsoDAR cyclotron: Its potential use for medical isotopes. The high currents of the IsoDAR cyclotron allow it to be used for the production of rare and low cross section isotopes \cite{waites2020potential,alonso2019medical}. Some of these isotopes have exciting applications in treatment as well as diagnostics in medicine. The use of \htp in IsoDAR allows it to be used for research and development for high power targets, as an \htp beam can be easily split among many target stations. The use of other ions which may be accelerated, particularly alpha particles, may also offer advantages to isotope production, without altering the physical cyclotron.

Therefore IsoDAR is not solely a tool for neutrino physics, but a more fundamental paradigm shift that pushes forward several fields simultaneously. By developing the IsoDAR cyclotron we not only have the ability to change physics, but to change lives. 

%% file: appa.tex
\chapter{Specific Contributions}

These papers within this work were done within a collaboration of approximately 30 people with a wide range of expertise. Here are my specific contributions:
\begin{itemize}
    \item In "High-current \htp beams from a filament-driven
multicusp ion source" \cite{winklehner2021high} I was a major part of the experimental setup, assembly, and commissioning. I wrote software and designed interfaces that expedited the mass spectrometry. I also simulated the beam dynamics during the upgrade process and executed sensitivity studies.

\item In "High intensity cyclotrons for neutrino physics" \cite{winklehner:nima} I completed the beam dynamics simulations for the LEBT matching to the RFQ, design of the LEBT, and commissioning of the ion source. 

\item In "What is the potential impact of the IsoDAR
cyclotron on radioisotope production: a
review"\cite{waites2020potential} and "Medical isotope production with the IsoDAR cyclotron" \cite{alonso2019medical}, I had completed yield estimates and established connections in the isotope industry in order to understand the needs of the community.

\item In "Axion-Like Particle Production at Beam Dump Experiments with Distinct Nuclear Excitation Lines" \cite{waites2022axion}I led and organized a group to model the axion yields in IsoDAR. I  used GEANT4 to find the nuclear photon yield, lepton yield, and nuclear excitations in the IsoDAR experiment. I then completed a phenomenological analysis and created a template to investigate processes which used photon coupling of the axion in the Yemilab detector.

\item In "Input Beam Matching and Beam Dynamics Design Optimizations of the IsoDAR RFQ Using Statistical and Machine Learning Techniques" \cite{koser2022input} I completed the machine learning analysis and optimization of the RFQ design.

\end{itemize}
\includepdf[pages={-}]{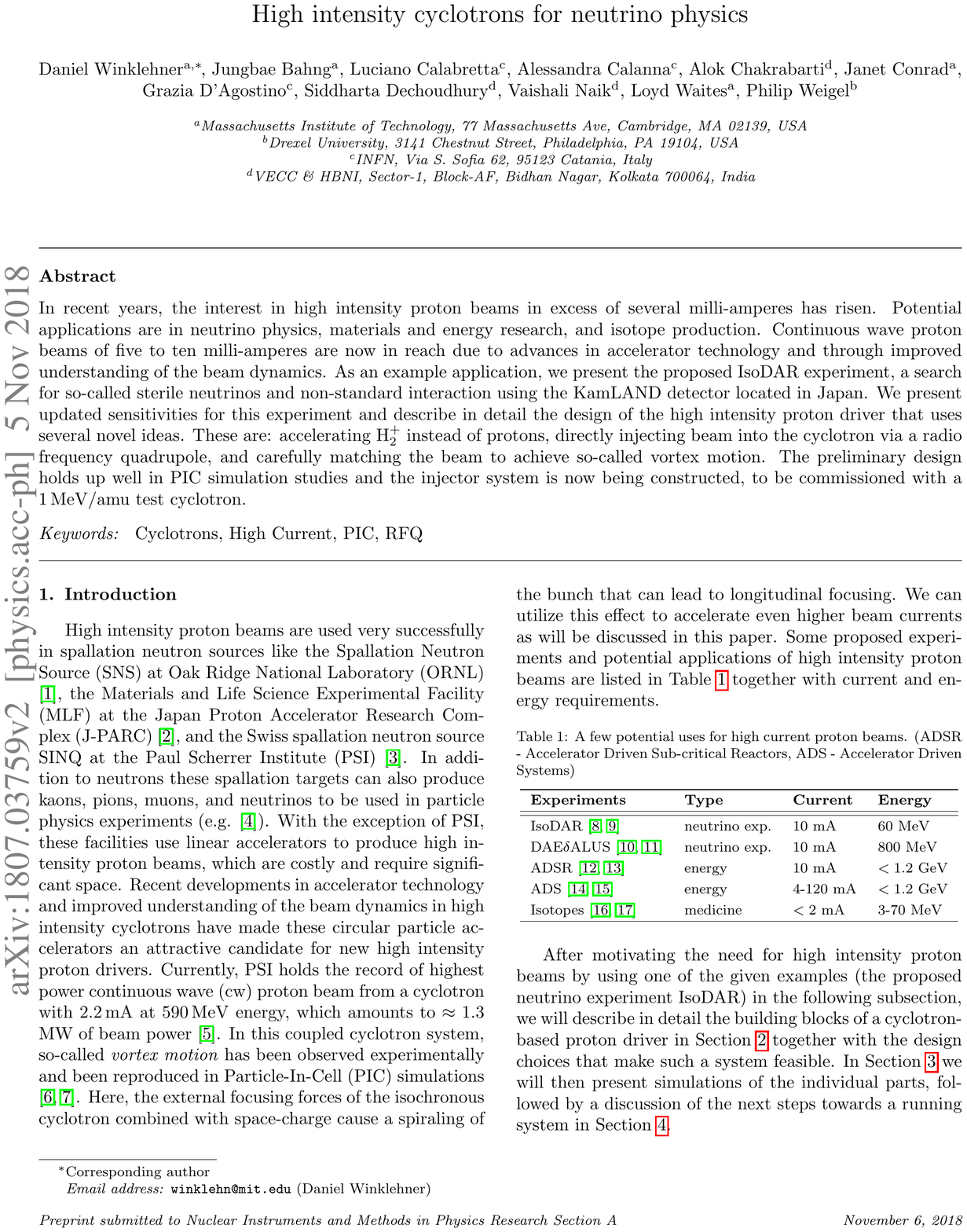}

\clearpage
\newpage

%% file: appb.tex
\chapter{Description of Ion Source}
\includepdf[pages={-}]{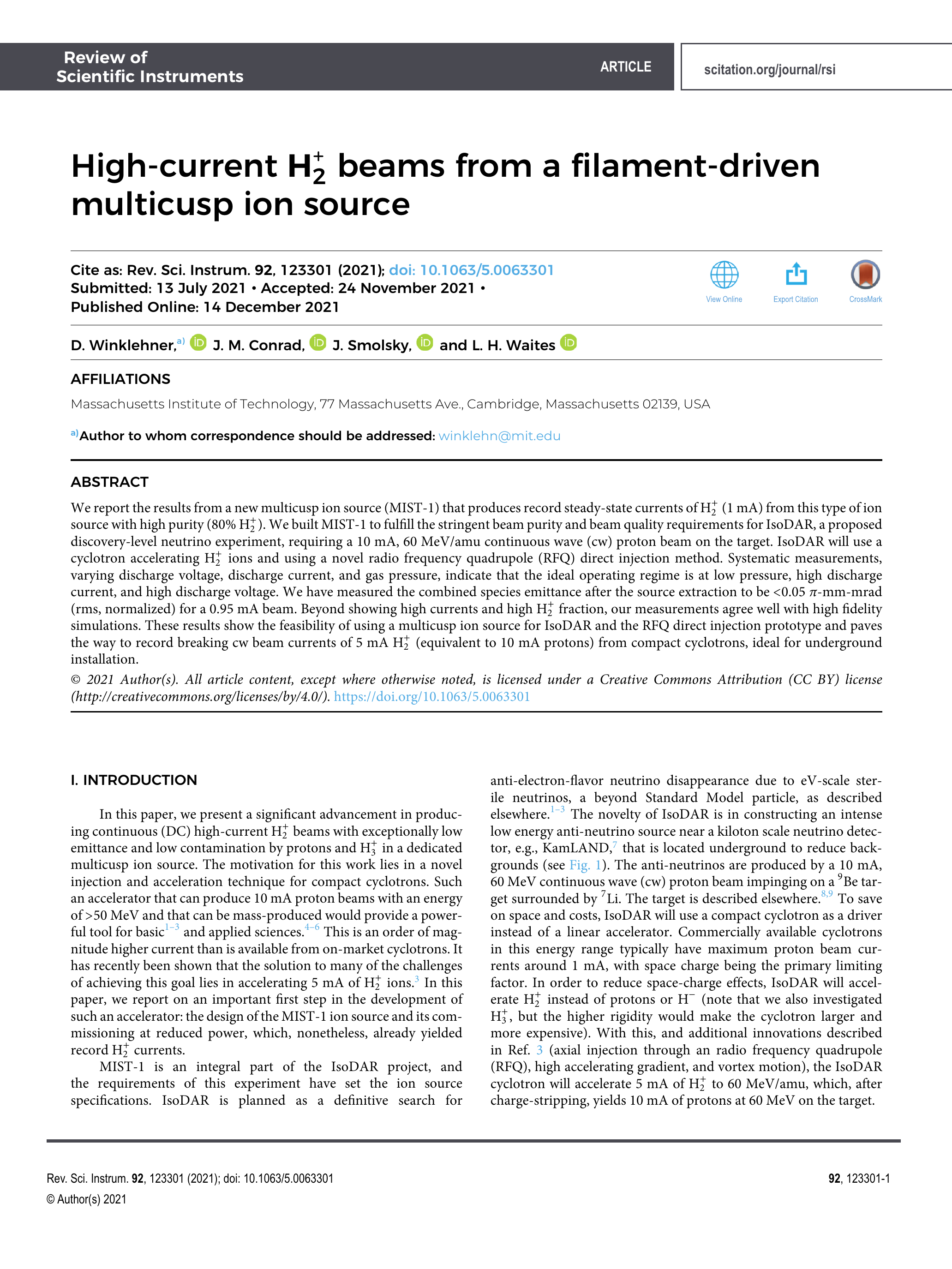}

\clearpage
\newpage

%% file: biblio.tex
\begin{singlespace}
\bibliography{main}
\bibliographystyle{plain}
\end{singlespace}